\documentclass[a4paper]{article}
\usepackage[T1]{fontenc}
\usepackage{lmodern}

\usepackage{geometry}
\usepackage{authblk}
\usepackage{setspace}
\usepackage[pdftex]{graphicx}
\graphicspath{{./figures/}}
\usepackage{subcaption}
\usepackage{amsmath, amssymb, mathrsfs}
\usepackage{float}
\usepackage{balance}
\usepackage{enumitem, array}
\usepackage{multirow, booktabs}
\usepackage{tabularx}
\usepackage{comment}
\usepackage{adjustbox}
\usepackage{booktabs}
\usepackage{caption}
\usepackage{xcolor}
\usepackage[table]{xcolor}
\usepackage{booktabs}
\usepackage{float}

\usepackage[colorlinks=true, citecolor=blue, linkcolor=blue, urlcolor=blue]{hyperref}

\usepackage[
    style=numeric-comp,
    citestyle=numeric,
    sorting=none,
    backend=biber
]{biblatex}
\usepackage{lineno}

\usepackage[labelfont=bf]{caption}

\DeclareUnicodeCharacter{0327}{\c{}}
\usepackage{longtable} 
\usepackage{booktabs}  
\usepackage{array}     

\title{Geological CO\textsubscript{2} storage assessment in emerging CCS regions: Review of sequestration potential, policy development, and socio-economic factors in Poland}

\author[1*]{Mohammad Nooraiepour}
\author[2]{Karol M. D\k{a}browski}
\author[1,3]{Mohammad Masoudi}
\author[2]{Szymon Kuczyński}
\author[4]{Zezhang Song}
\author[3]{Ane Elisabet Lothe}
\author[1]{Helge Hellevang}

\affil[1]{\small{Environmental Geosciences, Department of Geosciences, University of Oslo, P.O. Box 1047 Blindern, 0316 Oslo, Norway}}
\affil[2]{\small{Faculty of Drilling, Oil and Gas, AGH University of Krakow, al. Mickiewicza 30, 30-059, Krakow, Poland}}
\affil[3]{\small{SINTEF Industry, Applied Geoscience Department, 7465 Trondheim, Norway}}
\affil[4]{College of Geosciences, China University of Petroleum, Beijing, 102249, China}
\affil[*]{\small{Corresponding author: mohammad.nooraiepour@geo.uio.no}}

\begin{document}
\maketitle

\begin{abstract}
Emerging carbon capture and storage (CCS) markets face critical challenges in developing systematic methodologies to assess geological CO\textsubscript{2} storage potential under conditions of limited data availability, evolving regulatory frameworks, and nascent infrastructure development. This study establishes an assessment framework designed for lower-maturity CCS regions, using Poland as a representative case study to demonstrate methodology application and validate framework effectiveness. The framework integrates geological characterization, storage capacity assessment, regulatory analysis, and socio-economic evaluation through a structured approach adaptable to diverse global contexts. Poland's coal-reliant economy exemplifies the decarbonization challenges facing emerging CCS regions while meeting European Union climate mandates. The country's geological setting offers substantial sequestration opportunities across three major sedimentary regions. Through multidisciplinary analysis synthesizing scattered geological data, policy developments, CCUS value chain, and stakeholder perspectives, we systematically evaluate CO\textsubscript{2} storage potential. Onshore saline aquifers and depleted hydrocarbon fields provide significant storage capacity, while offshore Baltic Basin sites face logistical and environmental regulatory constraints. Current assessments encounter critical limitations, including sparse data, restricted research access, and inadequate industry-academia collaboration, preventing basin-scale analyses from advancing to higher storage readiness levels and undermining business decision-making reliability. This study contributes a replicable methodology extending beyond Poland to lower-maturity CCS regions worldwide. The framework provides decision-makers with systematic tools for storage assessment, policy development, and stakeholder engagement, supporting evidence-based CCS deployment strategies. Success in emerging markets requires coordinated advancement across technical characterization, regulatory clarity, infrastructure development, and public engagement, with transparent governance and inclusive community participation as critical enablers for sustainable CCS implementation.
\newline
\newline
\textbf{Keywords:} Carbon Capture and Storage (CCS), Geological CO\textsubscript{2} Storage, Storage Readiness Assessment, Saline Aquifers, Energy Transition, Stakeholder Engagement, Poland.
\end{abstract}


\section{Introduction}

Climate change, driven primarily by human-made carbon dioxide (CO\textsubscript{2}) emissions, demands urgent action to stabilize atmospheric concentrations and limit global warming \cite{kikstra2022ipcc}. These emissions contribute to rising global temperatures, erratic weather patterns, and significant disruptions to ecosystems worldwide \cite{kikstra2022ipcc, wheeler2013climate,tol2018economic,karl2009global}. Carbon capture and storage (CCS) has emerged as a pivotal technology for decarbonizing hard-to-abate sectors, offering the potential to capture CO\textsubscript{2} from industrial sources and store it securely in geological formations \cite{bui2018carbon,hellevang2015carbon,budinis2018assessment}.

This study conducts a geological CO\textsubscript{2} storage assessment in emerging CCS markets, using Poland as a case study to evaluate sequestration capacity in regions with developing infrastructure and policy frameworks. By synthesizing scattered scientific research, industrial reports, and evolving regulatory frameworks, this multidisciplinary review develops a systematic methodology to assess sequestration potential, providing a structured framework adaptable to lower-maturity global CCS regions. The analysis examines geological characteristics, technical and economic feasibility, policy development, and socio-economic factors that policymakers and industry stakeholders must navigate to formulate robust CCS strategies. This framework serves as a blueprint for high-level assessments, outlining standardized storage assessment protocols necessary to enhance large-scale implementation.

Within this global framework, the European Union (EU) has taken a leading role in climate policy. Initiatives such as the European Green Deal \cite{fetting2020european,claeys2019make} and the Fit for 55 package \cite{erbach2022fit,schlacke2022implementing} highlight the EU's commitment to achieving climate neutrality by the mid-century, with a mandated 55\% reduction in greenhouse gas emissions by 2030 compared to 1990 levels \cite{dupont2024three,wolf2021european}. These policies not only compel member states to adopt comprehensive decarbonization strategies.

\begin{figure}[h!]
    \centering 
    \includegraphics[width=0.75\textwidth]{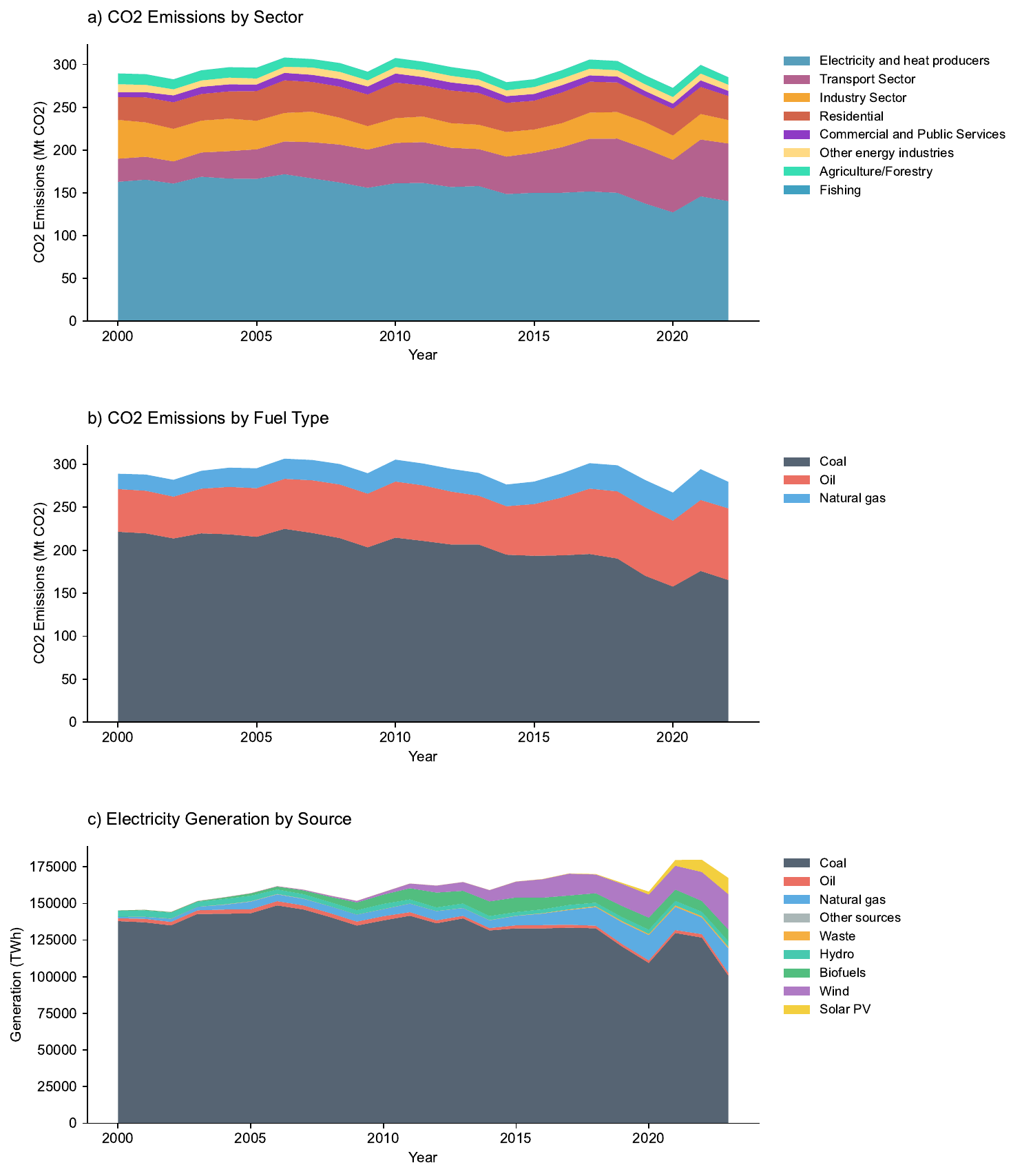}
    \caption{Temporal evolution of Poland's energy system and CO\textsubscript{2} emissions from 2000 to 2023.\\ 
    (a) CO\textsubscript{2} emissions by sector, dominated by electricity and heat production (contributing 49\% of total emissions in 2022), followed by transport (24\%) and industry sectors, illustrating the concentration of emissions in power generation and mobility. The transport sector shows steady growth, while residential emissions have declined due to efficiency improvements and fuel switching. 
    (b) CO\textsubscript{2} emissions by fuel type, highlighting Poland's persistent dependence on coal as the primary emission source (accounting for approximately 59\% of fuel combustion emissions in 2022), with oil maintaining a stable contribution primarily from transport, and natural gas showing a gradual increase as a transitional fuel in power generation and heating. 
    (c) Electricity generation by source, demonstrating Poland's ongoing energy transition from coal dominance (declining from 95\% in 2000 to 61\% in 2023) to a more diversified mix including rapid renewable energy expansion, particularly wind and solar photovoltaic systems that collectively reached 21\% of generation by 2023. The stacked visualization shows the scale of Poland's decarbonization challenge, with coal remaining the backbone of electricity supply despite policy commitments to reduce greenhouse gas emissions in line with EU climate targets (data sourced from \cite{IEA_Poland_Emissions_2023}).}
    \label{fig:poland_energy_stacked_trends}
\end{figure}

Poland presents a compelling case study as one of Europe's most carbon-intensive nations, historically dependent on fossil fuels, particularly coal \cite{sitnicki1991opportunities,kuchler2018down}. This legacy has resulted in substantial CO\textsubscript{2} emissions while creating economic and social ties to the fossil fuel industry \cite{surwillo2021public,radovanovic2017energy}. To understand the scale and structure of Poland's decarbonization challenge, Figure~\ref{fig:poland_energy_stacked_trends} shows the historical trends in CO\textsubscript{2} emissions by sector and fuel type, as well as the evolution of electricity generation sources over the past two decades.

Poland's emission profile reveals an evolving energy landscape (Figure~\ref{fig:poland_energy_stacked_trends}). Coal's share of total CO\textsubscript{2} emissions decreased from 77\% (221 Mt) in 2000 to 59\% (165 Mt) in 2022, while emissions from oil and natural gas increased. Electricity and heat production dominate Poland's emissions profile (49\% in 2022), followed by transport (23\%). Despite progress in renewable energy deployment—rising from negligible levels to 21\% of electricity generation by 2023 (Figure~\ref{fig:poland_energy_stacked_trends})—coal remains central to the energy mix, underscoring the urgent need for CCS integration in Poland's decarbonization strategy \cite{malec2022prospects,wojcik2021determinants}.

Economic and energy security considerations complicate the energy transition. Coal dependence has historically provided financial stability and energy self-sufficiency while fostering entrenched infrastructural and social connections to the fossil fuel industry \cite{vakulchuk2020renewable,karpinska2021will}. Transitioning to a low-carbon economy, therefore, requires not only technological innovation but also profound economic restructuring and policy reform. CCS emerges as a strategic enabler, offering a pathway to reduce emissions while maintaining energy reliability and economic competitiveness \cite{hewitt2019social}.

CCS in Poland has experienced alternating periods of intensive investigation and stagnation, largely reflecting the broader political and economic transitions in the country \cite{patricio2017region,uliasz2016perspectives,wesche2023ccus, nooraiepour2025norwegian}. Early research initiatives, primarily academic in nature, focused on exploring the feasibility of CO\textsubscript{2} capture and geological storage, laying the scientific foundation necessary for future applied projects \cite{budzianowski2011can,gabrus2021feasibility}. The advent of EU funding and international collaborations provided renewed momentum for Polish CCS initiatives. Collaborative projects, pilot studies, and feasibility assessments have since emerged, reflecting growing recognition of CCS as a critical component of the country's climate strategy \cite{budzianowski2011expansion}.

The Norwegian-Polish CCS network initiative exemplifies recent collaborative evolution, emphasizing the importance of assessing CO\textsubscript{2} storage options within Polish territory \cite{nooraiepour2025norwegian}. Geological storage of CO\textsubscript{2} encompasses both onshore and offshore candidates, each presenting distinct technical, economic, and environmental considerations \cite{delprat2015sitechar}. Assessment of domestic and transboundary CO\textsubscript{2} storage options requires advanced geological characterization techniques, simulation models, and risk assessment methodologies to develop robust evaluation frameworks for various storage candidates \cite{brunsting2015ccs,radoslaw2009co2,delprat2015sitechar}.

The March 2025 memorandum of understanding between Poland's ORLEN and Norway's Equinor represents a significant milestone in bilateral CCS collaboration \cite{orlen2025ccs}. This partnership aims to identify potential CO\textsubscript{2} storage sites both onshore and within the Polish sector of the Baltic Sea \cite{offshore2025ccs,reuters2025ccs}. ORLEN has established a strategic target of capturing, transporting, and storing 4 million metric tons of CO\textsubscript{2} annually by 2035, utilizing this capacity for its petrochemical and refining operations while offering remaining capacity as a service to other industries \cite{orlen2025ccs, reuters2025ccs}. Equinor contributes extensive CCS experience, having pioneered CO\textsubscript{2} storage at the offshore Sleipner field in 1996 and leading several large-scale CCS projects, including the Northern Lights project—the first cross-border CCS initiative providing CO\textsubscript{2} storage as a commercial service \cite{nooraiepour2025norwegian}.

Baltic Sea carbon storage faces significant environmental, technical, and regulatory constraints beyond geological suitability. The shallow, brackish environment and ecological sensitivity raise concerns about CO\textsubscript{2} leakage. The Helsinki Convention's prohibition on "dumping" creates legal complexity that may extend to CO\textsubscript{2} storage, contrasting with the accommodating London Protocol framework \cite{dixon2021exporting}. Competing offshore activities—shipping, fishing, and renewable energy installations—complicate site selection, while the region's distinctive geology requires costly adaptation of North Sea storage techniques and dedicated measurement, monitoring and verification (MMV) systems, potentially constraining economic viability. Regional regulatory coordination presents additional complexity for Baltic Sea CCS projects. The Helsinki Convention requires consensus among all Baltic Sea states for activities that could affect marine environments, creating potential delays in project approvals. Additionally, the intersection of national energy security concerns with international climate objectives may influence the pace and scope of regional cooperation on cross-border storage initiatives.

This study addresses critical knowledge gaps in CCS assessment methodologies for emerging markets by developing a comprehensive framework that integrates geological, regulatory, and socio-economic analyses. Poland serves as an exemplary case study due to its substantial storage potential, evolving policy landscape, and position as a representative emerging CCS market in Central and Eastern Europe. The research synthesizes scattered data sources and diverse subsurface resources to critically evaluate CO\textsubscript{2} storage potential in both onshore saline aquifers and depleted hydrocarbon fields, as well as offshore Baltic Basin sites. By addressing data scarcity, restricted research access, and limited industry-academia collaboration, the study develops systematic methodologies to assess sequestration capacity with confidence intervals appropriate for business decision-making.

Key contributions include: (1) a structured assessment framework adaptable to global emerging CCS regions; (2) comprehensive evaluation of Poland's storage potential across multiple geological formations; (3) analysis of policy evolution and regulatory frameworks essential for CCS deployment; (4) identification of infrastructure requirements and value chain integration challenges; and (5) strategies for transparent governance and inclusive community engagement to foster CCS acceptance. Positioned at the intersection of Poland's coal-reliant economic heritage and environmental imperatives, this study advances understanding of geological sequestration while providing actionable insights for resilient and environmentally responsible strategies. The generalized framework enriches knowledge that guides CCS implementation across Europe and other emerging markets, supporting a balance between energy security and climate goals.

This review is organized as follows: Section 2 traces the evolution of CCS/CCUS policy in Europe and Poland, analyzing regulatory developments and funding mechanisms. Section 3 presents a thorough methodological framework for high-level storage capacity assessments in less-developed and emerging regions, including the resource-reserve pyramid and Storage Readiness Levels. To set an example, Section 4 examines Poland's geological setting and major sedimentary basins relevant to CO\textsubscript{2} storage. Section 5 evaluates the storage potential of specific sedimentary basins and provides capacity estimates. Section 6 assesses carbon mineralization potential in mafic and ultramafic rocks. Section 7 synthesizes findings through discussion of geological storage status, value chain integration, infrastructure and regulatory considerations, monitoring frameworks, and public engagement considerations.

\section{Evolution of Policy and Regulatory Frameworks}

The European Union (EU) has established a policy framework to combat climate change, promoting CCS/CCUS as critical technologies for decarbonization. The 2009 EU CCS Directive 2009/31/EC set standards for safe geological CO\textsubscript{2} storage, covering site selection, monitoring, and liability \cite{EU_CCS_Directive_2009}. The 2018 Directive 2018/2001 on renewable energy supported synthetic fuels from captured CO\textsubscript{2}, while the EU Emissions Trading System (ETS) Directive exempted permanently stored CO\textsubscript{2} from emission allowances \cite{grubb2006allocation}. The 2013 EU 2030 Climate and Energy Framework targeted a 20\% emissions reduction by 2030 relative to 1990, encouraging CCS research \cite{kulovesi2020assessing,}. The 2019 European Green Deal aimed for net-zero emissions by 2050, prioritizing CCS for hard-to-abate sectors \cite{EUGreenDeal2019}. The 2020 EU 2030 Climate Target Plan raised the emissions reduction goal to 55\% by 2030, emphasizing industrial decarbonization \cite{rivas2021towards}.

The EU Taxonomy classifies CCS and CCU as significant contributors to climate change mitigation, enabling access to EU funding \cite{lucarelli2020classification,schutze2024eu}. The 2022 Corporate Sustainability Reporting Directive (CSRD) mandates environmental, social, and governance (ESG) reporting for large companies from 2024 and listed Small and medium enterprises from 2026, enhancing transparency for CCS/CCU investments \cite{baumuller2021moving,odobavsa2023expected}. The 2021 Communication COM/2021/800 on sustainable carbon cycles set targets for absorbing 310 Mt CO\textsubscript{2}eq via land and biomass and 5 Mt CO\textsubscript{2}eq from industrial facilities by 2030 \cite{wolf2022sustainable,tomkins2019evaluating}. In 2022, the EU proposed a voluntary carbon removal certification framework for industrial and biomass-based CO\textsubscript{2} removal, such as carbon farming \cite{gunther2024carbon,brad2023carbon}. The 2021 Carbon Border Adjustment Mechanism (CBAM) incentivizes CCS in carbon-intensive sectors by imposing import carbon costs \cite{bellora2023eu,zhong2024carbon}. The 2023 Net-Zero Industry Act (NZIA) designated CCS as a strategic technology, targeting 50 Mt CO\textsubscript{2} injection annually by 2030, streamlining permitting within 18 months, and mandating Member States to identify storage sites with private sector involvement, particularly from oil and gas \cite{tagliapietra2023rebooting,guillaume2025implementing}. The 2024 EU Industrial Carbon Management Strategy projected 80 Mt storage by 2030, 300 Mt by 2040, and 550 Mt by 2050, requiring €12B and 7,500 km of CO\textsubscript{2} pipelines by 2030, and €16B and 19,000 km by 2040 \cite{maatta2025resistance,maatta2024analysis}. In July 2024, updated non-binding Guidance Documents for CCS permitting were released to harmonize licensing and attract investment \cite{arguello2024transboundary,dbouk2024review}. The EU Innovation Fund, launched in 2018, allocated €4.8B by 2024 for CCS/CCU projects \cite{bachtrogler2020eu,rienks2023eu}. The EU ETS Phase IV (2021--2030) targets a 62\% CO\textsubscript{2} reduction by 2030 compared to 2005, increasing carbon prices for high-emission industries \cite{zaklan2021eu}.

\begin{figure}
    \centering
    \includegraphics[width=1.00\textwidth]{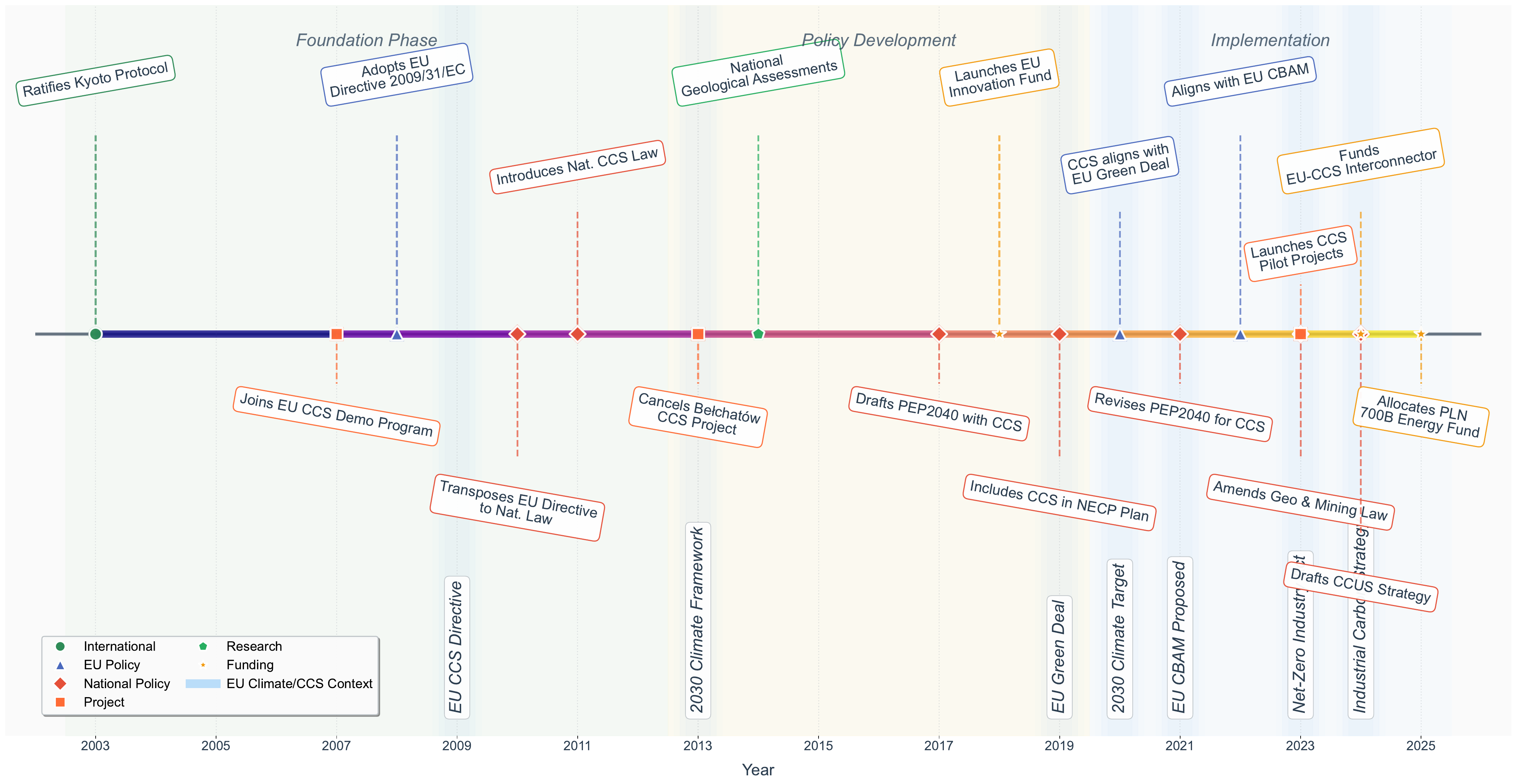}
    \caption{Timeline of Poland's CCS and CCUS policy evolution (2003-2025) in the context of EU regulatory development. Major milestones are categorized by type: international agreements (Kyoto Protocol), EU policies (CCS Directive 2009/31/EC, Green Deal, Net-Zero Industry Act), national legislation (CCS Act 2011, NECP, PEP2040), industrial projects, research initiatives, and funding mechanisms. EU climate directives are represented by shaded background bands, with a temporal progression across three distinct phases: foundation, policy development, and implementation.}
    \label{fig:F3_ccstimeline}
\end{figure}

Poland’s CCS/CCUS framework has evolved over the past two decades, transitioning from legal foundations to practical strategies, culminating in pilot and eventually large-scale projects in the years to come, driven by EU imperatives and national decarbonization goals. This evolution, marked by legal, strategic, funding, and project milestones, is illustrated in Figure~\ref{fig:F3_ccstimeline}.

\paragraph{2008--2014: Establishing Legal and Geological Foundations}
Poland’s CCS/CCUS journey began with its 2003 ratification of the Kyoto Protocol, which emphasized innovative emission reduction technologies \cite{szymczyk2006problemy}. In 2007, Poland joined the EU CCS demonstration program, launching the Bełchatów CCS project for its largest coal plant, funded by the European Energy Programme for Recovery (EEPR). The project was abandoned in 2013 due to local opposition, economic concerns, and legal uncertainties over offshore storage \cite{gibbins2008preparing, pyc2008aspekty}. In 2008, the EU CCS Directive 2009/31/EC prompted Poland to transpose its provisions into national law by 2010 through amendments to the Geological and Mining Law \cite{EU_CCS_Directive_2009}. The 2011 CCS Act clarified exploration and storage licensing procedures and introduced environmental impact assessment requirements. By 2014, geological surveys by the Polish Geological Institute and international partners estimated significant CO\textsubscript{2} storage capacity in deep saline aquifers and in the Baltic Sea basin, integrating CCS into the National Program for the Reduction of Greenhouse Gases \cite{wojcicki2014assessment, anthonsen2014screening, DzU2015_2273}. These measures established the legal and geological foundation for CCS/CCUS development, but subsequent disruptions significantly hindered progress.

\paragraph{2015--2019: Planning and Initial Commitments}
Since 2015, Poland has been gradually and slowly integrating CCS/CCUS into its energy transition framework. The 2017 draft of the Polish Energy Policy until 2040 (PEP2040) identified CCS as a decarbonization tool for coal regions and heavy industry, aligning with its pillars of Just Transition, Zero-Emission Energy System, and Good Air Quality \cite{tajdus2020risks}. In 2019, the National Energy and Climate Plan (NECP) 2021--2030, prioritized research into CO\textsubscript{2} transport/storage and CCU pathways (e.g., CO\textsubscript{2}-to-methane and methanol production) as part of the Strategic Directions for Energy Innovation Development, but deferred binding CCS deployment targets in favor of renewables and natural gas \cite{wojtkowska2021eu, pluta2019review}. Concurrently, Poland joined the Clean Energy for EU Islands Initiative, securing funding for feasibility studies to develop Baltic region CCS hubs \cite{kotzebue2020eu,winter2023eu}. These initiatives marked the initial commitments despite a primary focus on other energy sources.

\paragraph{2020--2025: Advancing National CCUS Strategy}
Since 2020, Poland has accelerated its regulatory reforms for CCS/CCUS, funding, and projects, aligning with EU climate goals. In 2020, Poland adopted the EU Green Deal’s 50 Mt CO\textsubscript{2} storage target by 2030 \cite{EUGreenDeal2019}. The 2021 PEP2040 emphasized CCS for coal region transformation, supported by a social agreement to develop CO\textsubscript{2} transport infrastructure and underground storage from 2023--2029. PEP2040 also linked CCS to low-carbon hydrogen production via steam methane reforming, supporting the 2021 Polish Hydrogen Strategy, and highlighted its role in electromobility and alternative fuels, including methanol production \cite{panstwowych2021transformacja, dragan2021polish}. Feasibility studies for CCS in cement, steel, and chemical industries began to emerge. In October 2023, amendments to the Geological and Mining Law and Energy Law lifted restrictions, enabling (at least in theory) industrial-scale CCUS deployment beyond demonstration projects. These amendments simplified licensing, exempted small installations (<100 kt CO\textsubscript{2} total storage), permitted direct CO\textsubscript{2} transport, and introduced CO\textsubscript{2}-enhanced oil recovery (EOR) to improve profitability. Investors were exempted from EU ETS emission allowances for captured and stored CO\textsubscript{2}, reducing costs. The 2023 National Recovery and Resilience Plan (KPO) allocated €2.5B for CCS/CCU research and pilot projects \cite{zeitlin2024national,florczak2022national}. Moreover, the draft of the national CCUS strategy and the pilot concept of the Polish CCUS Cluster were materialized. The Energy Transformation Fund will support onshore and offshore further developments, aligning with Poland’s EU Council Presidency, where CCS is positioned to influence EU policy \cite{szafran2023finansowanie, kawecka2022financing}.

Key projects include the Go4ECOPlanet project (2023–2033), funded with €228M from the EU Innovation Fund, targeting 100\% CO\textsubscript{2} capture from the Kujawy cement plant, with liquefied CO\textsubscript{2} transported to North Sea storage sites. The ORLEN–Equinor Memorandum of Understanding aims for 4 Mt/year CO\textsubscript{2} storage by 2035 through joint assessment of onshore and Baltic Sea sites \cite{orlen2025ccs}. The Poland--EU CCS Interconnector, a project of common interest, received €2.5M from the Connecting Europe Facility in 2024 for feasibility studies. Additional financing comes from Horizon Europe, the European Bank for Reconstruction and Development, and the European Investment Bank \cite{kawecka2022financing,dembicka2023structural}.

The cement industry, accounting for 3.8\% of Poland’s CO\textsubscript{2} emissions, faces pressure to adopt CCS due to the EU ETS phase-out of free CO\textsubscript{2} allowances by 2034, with reductions starting in 2026. The Kujawy CCS installation, planned for 2027, aims to capture 2.5 Mt CO\textsubscript{2} annually by 2030 and potentially 100\% of emissions by 2040, though high costs (0.5--1.5B PLN per installation) remain a challenge.

\section{Methodologies for Storage Capacity Assessments}
\label{sec:SRLmethodology}

This section establishes the methodological framework for systematic CO\textsubscript{2} assessment in emerging CCS regions, demonstrated through Poland's geological evaluation. The approach integrates internationally recognized assessment protocols—the techno-economic resource-reserve pyramid \cite{kopp2009investigations, bradshaw2007co2} and Storage Readiness Level (SRL) classification \cite{Akhurst2019storageReadiness}—with regional geological data synthesis and multi-scale modeling techniques. Poland's assessment draws from major European subsurface characterization initiatives, including the EU GeoCapacity \cite{vangkilde2009assessing}, CO\textsubscript{2}STOR \cite{chadwick2008best}, and CCUS ZEN projects \cite{Ringstad2023CCUSZEN}, providing both methodological validation and contextual grounding for the broader framework development.

\subsection{Resource-Reserve Pyramid and Storage Capacity Tiers}
\label{subsec:reserve-pyramid}

\begin{figure}[h]
    \centering
   \includegraphics[width=0.75\textwidth]{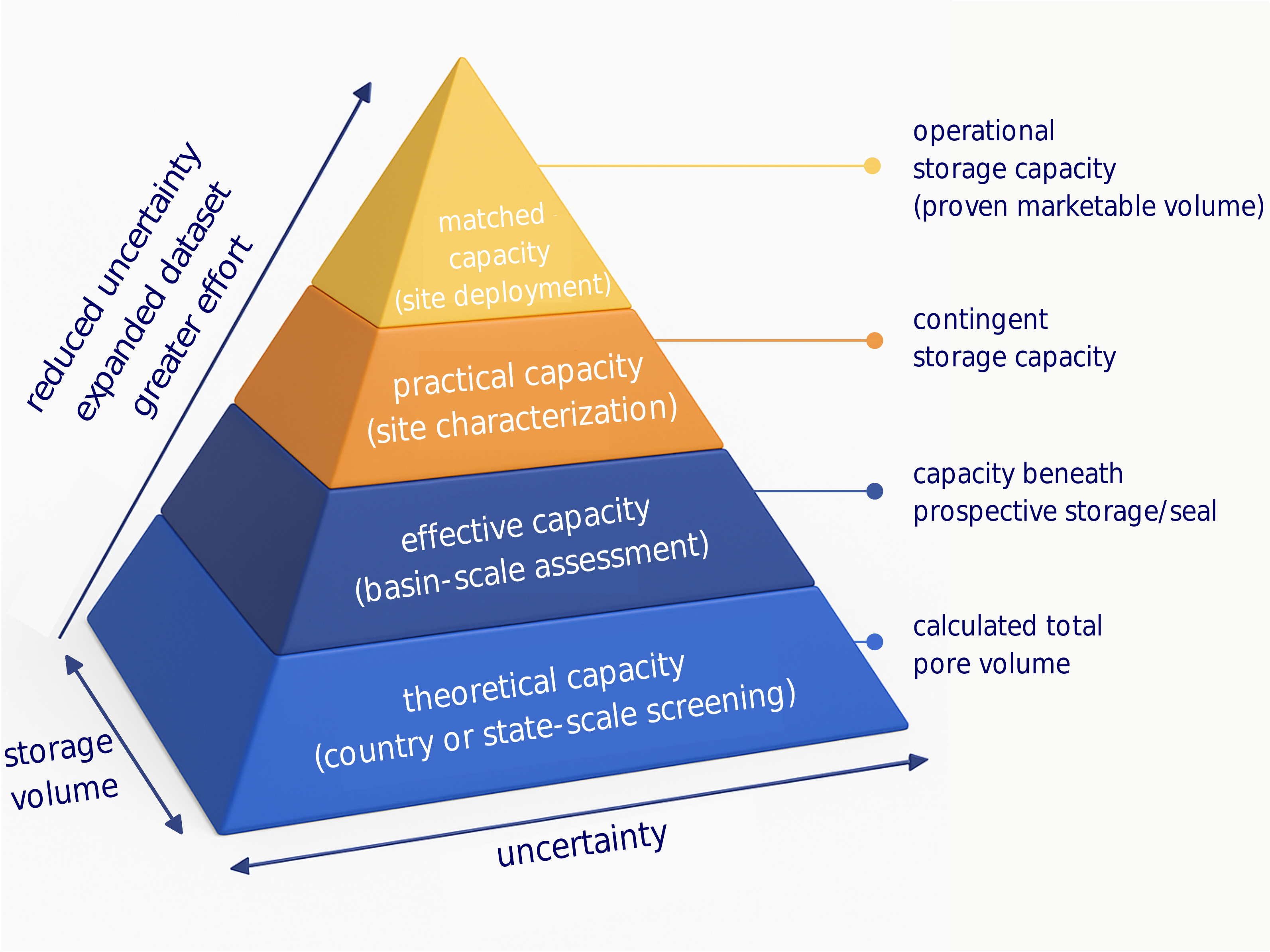}
   \caption{Techno-economic pyramid of CO\textsubscript{2} storage capacity progression from theoretical geological potential to market-ready storage volumes. The four tiers represent: theoretical (maximum geological potential), effective (geologically suitable basin-scale sites), practical (characterized sites with infrastructure assessment), and matched (operationally proven volumes). Higher tiers require increased data, effort, and certainty while available storage volumes decrease.}
    \label{fig:F4_pyramid}
\end{figure}

The assessment of CO\textsubscript{2} storage capacity is structured within the techno-economic resource-reserve pyramid, a widely adopted framework in CCS research \cite{kopp2009investigations, bachu2005co2, bradshaw2007co2}. This hierarchical model, illustrated in Figure~\ref{fig:F4_pyramid}, categorizes storage potential into four progressive tiers, each imposing additional constraints to refine estimates from theoretical maxima to operationally viable capacities.

\subsubsection{Theoretical Capacity}
Theoretical capacity represents the upper physical limit of CO\textsubscript{2} storage, assuming all pore space within a geological formation is fully accessible for CO\textsubscript{2} storage (Fig.~\ref{fig:F4_pyramid}), either as a free phase or dissolved in formation fluids. It is calculated using a volumetric approach \cite{bachu2007co2}:
\begin{equation}
    C_{\text{theo}} = A \times h \times \phi \times \rho_{\text{CO}_2} \times S_{\text{CO}_2},
    \label{eq:theoretical}
\end{equation}
where \( A \) (m\textsuperscript{2}) is the formation area, \( h \) (m) is the gross thickness, \( \phi \) (dimensionless) is the porosity, \( \rho_{\text{CO}_2} \) (kg/m\textsuperscript{3}) is the CO\textsubscript{2} density at reservoir conditions, and \( S_{\text{CO}_2} \) (dimensionless) is a storage efficiency factor reflecting total pore accessibility. For saline aquifers, \( S_{\text{CO}_2} \) may approach unity in idealized scenarios, while in hydrocarbon reservoirs, it is constrained by original fluid volumes \cite{goodman2011us}.

Theoretical estimates in Polish assessments rely on broad geological mapping but lack site-specific refinement. Assumptions include uniform porosity and full accessibility, which yield overly optimistic estimates but overlook geological complexities. 

\subsubsection{Effective Capacity}
Effective capacity refines theoretical capacity by applying geological and engineering constraints, such as depth, salinity, porosity-permeability cutoffs, and formation heterogeneity. It is a subset of theoretical capacity, adjusted via efficiency factors (\( E_{\text{saline}} \) for aquifers, \( E_{\text{hydrocarbon}} \) for depleted reservoirs) that account for irreducible water saturation, pressure limits, and injectivity. The formulation for saline aquifers, adapted from USDOE methodologies \cite{goodman2011us}, is:
\begin{equation}
    C_{\text{eff}} = A \times h_g \times \phi \times N_{TG} \times \rho_{\text{CO}_2} \times E_{\text{saline}},
    \label{eq:effective-aquifer}
\end{equation}
where \( h_g \) (m) is the gross thickness, \( N_{TG} \) (dimensionless) is the net-to-gross ratio, and \( E_{\text{saline}} \) (typically 1–20\%) reflects usable pore volume and displacement efficiency. 

For hydrocarbon reservoirs, effective capacity leverages displaced hydrocarbon volumes \cite{bachu2005co2}:
\begin{equation}
    C_{\text{eff, oil}} = C_e \times \rho_{\text{CO}_2} \times RF \times OOIP \times B_f - (V_{iw} - V_{pw}),
    \label{eq:effective-oil}
\end{equation}
\begin{equation}
    C_{\text{eff, gas}} = C_e \times \rho_{\text{CO}_2} \times RF \times OGIP \times \left( \frac{p_{\text{res}} T_{\text{sc}} Z_{\text{sc}}}{p_{\text{sc}} T_{\text{res}} Z_{\text{res}}} \right),
    \label{eq:effective-gas}
\end{equation}
where \( RF \) is the recovery factor, \( OOIP \) and \( OGIP \) are original oil and gas in place, \( B_f \) is the formation volume factor, and \( V_{iw} \), \( V_{pw} \) are injected/produced water volumes (often omitted due to data scarcity). \( C_e \) (e.g., 0.25–0.48 for oil, 0.63–0.87 for gas) adjusts for mobility and aquifer effects \cite{bachu2005co2}.

\subsubsection{Practical Capacity}
Practical capacity narrows effective capacity by incorporating technical, economic, and regulatory constraints, such as infrastructure availability, injection costs, and legal frameworks. It requires operational data from pilot projects, which remain limited in Poland post-Bełchatów’s cancellation \cite{nooraiepour2025norwegian}. Estimates decrease significantly at this stage due to site-specific challenges (Fig.~\ref{fig:F4_pyramid}), such as legacy well integrity and restrictions in the Baltic Sea, as mentioned in the Introduction. Limited pilot data significantly restricts Poland’s practical capacity estimates. 

\subsubsection{Matched Capacity}
Matched capacity aligns CO\textsubscript{2} sources (e.g., industrial emitters) with storage sites, optimizing injectivity, capacity, and proximity. Poland’s ORLEN-Equinor collaboration (2025) may ultimately demonstrate this stage, though detailed assessments remain pending or confidential, awaiting future realization.

\subsection{Storage Readiness Level (SRL)}
\label{subsec:storage readiness}

The Storage Readiness Level (SRL) framework is a standardized tool designed to assess the maturity of geological sites for CO\textsubscript{2} storage \cite{Akhurst2019storageReadiness}. Inspired by the Technology Readiness Level (TRL) system, SRL evaluates storage sites from initial identification to operational readiness, encompassing technical, regulatory, and operational milestones. 

\begin{figure}[h!]
    \centering    
    \includegraphics[width=\textwidth]{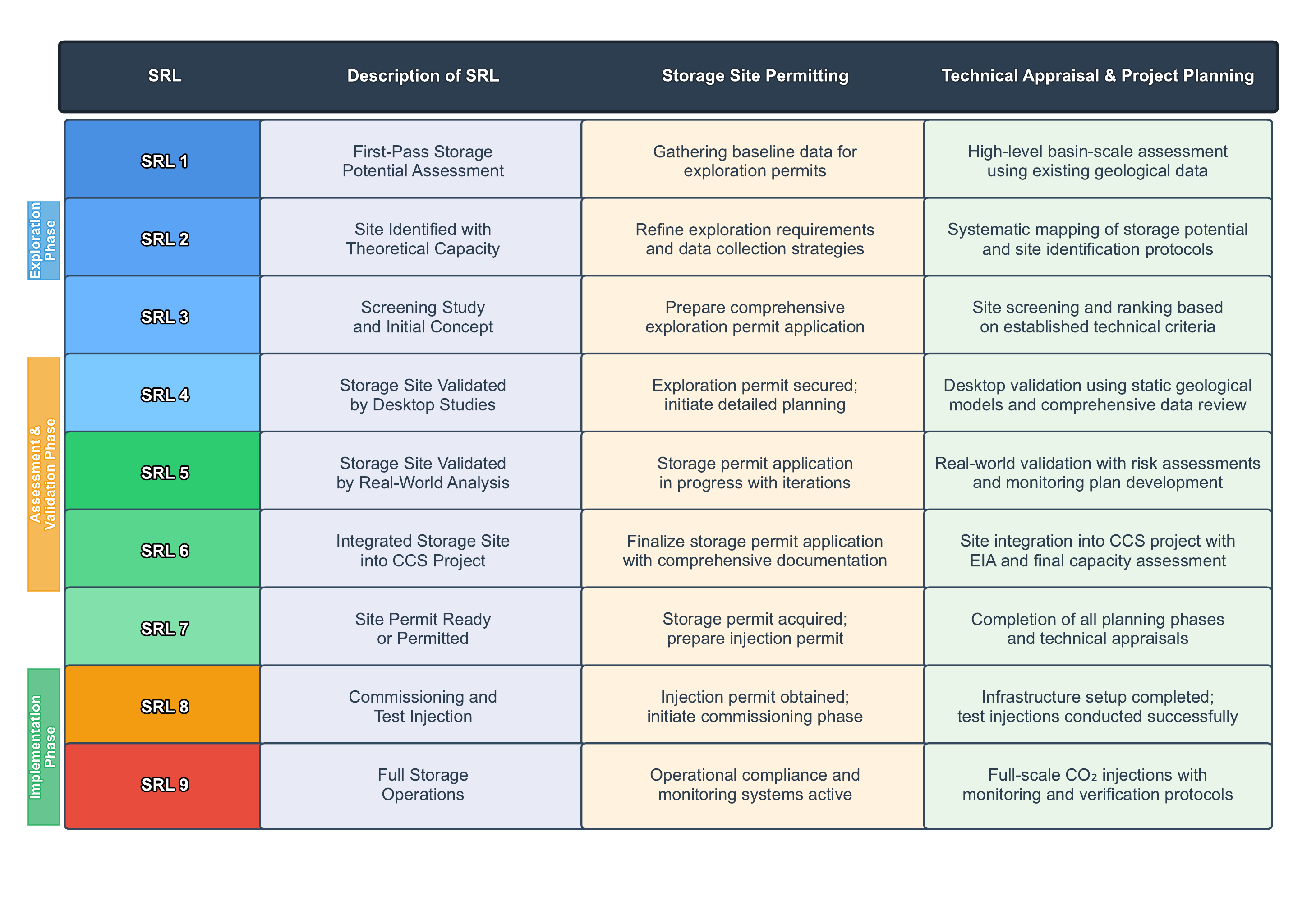}
    \caption{Storage Readiness Level (SRL) framework for CO\textsubscript{2} storage site development, showing the systematic progression from initial assessment (SRL 1) through full operational status (SRL 9). The framework integrates three key dimensions: technical appraisal and project planning, storage site permitting requirements, and milestone descriptions for each readiness level. Color-coded phases distinguish exploration (SRL 1-2), assessment and validation (SRL 3-6), and implementation stages (SRL 7-9).}
    \label{fig:F5_srl}
\end{figure}

Figure \ref{fig:F5_srl} shows that the SRL framework is derived by integrating three core pillars: technical appraisal, permitting, and operational planning. Each level requires specific evidence to justify progression. Technical appraisal involves detailed geological studies, including seismic surveys, well data analysis, and reservoir modeling, to estimate storage capacity and ensure containment integrity through mechanisms such as structural trapping and residual trapping. Permitting evaluates compliance with national and international regulations, which mandate environmental impact assessments and public engagement. Operational planning addresses logistical requirements, including transport infrastructure and monitoring systems for long-term site management. As Figure \ref{fig:F5_srl} outlines, the SRL scale ranges from SRL 1 (conceptual site identification) to SRL 9 (fully operational storage site), with intermediate levels marking milestones like feasibility studies (SRL 4--6) and pilot injections (SRL 7--8). The framework accommodates site-specific challenges and regulatory variations by prioritizing sites with high readiness and reducing uncertainties in storage capacity and safety \cite{Akhurst2019storageReadiness, Ringstad2023CCUSZEN}.

The SRL framework expands on the techno-economic resource-reserve pyramid assessment (Fig. \ref{fig:F4_pyramid}) and provides indicators for evaluating the suitability of geological sites for CO\textsubscript{2} storage (Fig. \ref{fig:F5_srl}). SSRL 1–3, as applied in the following sections, are defined as follows: SRL 1 involves a preliminary, state- or country-scale assessment to estimate theoretical storage capacity and identify geological characteristics, such as suitable lithology and structural traps. At SRL 2, sites with theoretical capacity are systematically mapped, with assessments based on geological data and theoretical calculations of storage potential, laying the groundwork for more targeted evaluations. SRL 3 advances to a detailed screening study, focusing on individual storage sites and developing an initial project concept by integrating geological, technical, economic, and geographical criteria (Fig. \ref{fig:F5_srl}). The EU-funded CCUS ZEN project recently reevaluated prior assessments of Polish subsurface storage potential \cite{wojcicki2014assessment} and classified a list of identified CO\textsubscript{2} storage sites in Poland at SRL 2--3, indicating early-stage geological and technical suitability \cite{Ringstad2023CCUSZEN}.

\subsection{Assessment Framework for Storage Potential}
\label{subsec:poland-framework}

In the context of CO\textsubscript{2} geological storage, a prospect unit is a critical classification for identifying and developing viable storage sites, encompassing several key geological and technical concepts. Reservoir formations, characterized by favorable reservoir properties such as high porosity and permeability, form the foundation for effective CO\textsubscript{2} storage, ensuring long-term retention of carbon dioxide. Within these formations, storage units are defined as coherent, mappable subsurface bodies of reservoir rock located at sufficient depths, sharing similar geological characteristics that enable secure CO\textsubscript{2} containment with minimal leakage risk. Daughter units, which are specific structural and stratigraphic traps within a storage unit, include depleted hydrocarbon fields that leverage their established geological structures for safe storage. A prospect unit, the most refined classification, is a daughter unit evaluated as commercially viable and bankable for CO\textsubscript{2} storage, marking its potential for development in CCS projects.

\subsubsection{Reservoir Characterization and Static Modeling}
\label{subsec:reservoir-characterization}

High-resolution 3D geological models are foundational for estimating static storage capacity, integrating well logs, core samples, seismic surveys, and other geophysical measurements. These models delineate reservoir geometry and quantify petrophysical properties (e.g., porosity, permeability, and fluid saturation) that are essential for static and dynamic simulations. Advanced workflows employ seismic inversion and geostatistical techniques to enhance model fidelity. Static capacity is calculated volumetrically \cite{bachu2007co2}:
\begin{equation}
    V_{\text{CO}_2} = A \times h \times \phi \times \rho_{\text{CO}_2} \times E,
    \label{eq:static}
\end{equation}
where \( A \) (m\textsuperscript{2}) is the effective area, \( h \) (m) is the net thickness, \( \phi \) (dimensionless) is the porosity, \( \rho_{\text{CO}_2} \) (kg/m\textsuperscript{3}) is the supercritical CO\textsubscript{2} density, and \( E \) (dimensionless) is an efficiency factor (e.g., 1–20\% for aquifers, 50–80\% for depleted reservoirs) accounting for sweep efficiency, heterogeneity, and trapping mechanisms \cite{gorecki2009development,}. Sensitivity analyses, using Monte Carlo simulations with triangular distributions (P10, P50, P90), can further refine uncertainty bounds.

\subsubsection{Dynamic Flow Modeling}
\label{subsec:dynamic-modeling}

Dynamic flow modeling simulates CO\textsubscript{2} injection behavior, integrating multiphase fluid dynamics, mass transport, and geomechanical interactions. Guided by conservation laws, these models accurately predict plume migration, pressure evolution, and caprock integrity, which are crucial for effective operational planning and risk management. The core equations are \cite{bachu2007co2,ajayi2019review}:

\begin{align}
    \text{Mass Conservation (Multiphase Flow):} & \quad \frac{\partial}{\partial t} (\phi \rho S) + \nabla \cdot (\rho \mathbf{v}) = Q, \label{eq:mass} \\
    \text{Momentum Conservation (Darcy’s Law):} & \quad \mathbf{v}_\alpha = -\frac{k k_{r\alpha}}{\mu_\alpha} (\nabla p_\alpha - \rho_\alpha \mathbf{g}), \label{eq:momentum} \\
    \text{Energy Conservation (Heat Transport):} & \quad \rho c_p \frac{\partial T}{\partial t} + \nabla \cdot (-k_T \nabla T) = H_{\text{source}}, \label{eq:energy}
\end{align}

where \( \mathbf{v}_\alpha \) (m/s) is the Darcy velocity of phase \( \alpha \), \( k \) (m\textsuperscript{2}) is absolute permeability, \( k_{r\alpha} \) (dimensionless) is relative permeability, \( \mu_\alpha \) (Pa·s) and \( \rho \) (kg/m\textsuperscript{3}) are viscosity and density, \( \phi \) (dimensionless) is porosity, \( S \) (dimensionless) is saturation, \( Q \) (kg/m\textsuperscript{3}/s) is a source/sink term, \( \mathbf{g} \) (m/s\textsuperscript{2}) is gravitational acceleration, \( c_p \) (J/kg·K) is heat capacity, \( k_T \) (W/m·K) is thermal conductivity, and \( H_{\text{source}} \) is a heat source term.

\subsubsection{Risk-Weighted Capacity and Uncertainty Analysis}
\label{subsec:risk-analysis}

A risk-weighted approach bridges static and dynamic assessments, adjusting dynamic capacity (\( C_{\text{dyn}} \)) with geological and operational risks \cite{bradshaw2003australia}:
\begin{equation}
    C_{\text{eff}} = C_{\text{dyn}} \prod_{i=1}^{n} (1 - R_i),
    \label{eq:risk-weighted}
\end{equation}
where \( R_i \) represents risks (e.g., fault reactivation, well leakage, maximum allowable pressure), quantified via techniques such as probabilistic and Monte Carlo-type methods. An integrated framework could also employ multi-criteria decision analysis (MCDA), balancing technical feasibility, economic viability, and environmental safety.

\subsection{Data Compilation and Regional Constraints}
\label{subsec:data-compilation}

The evaluation of CO\textsubscript{2} storage potential in emerging regions hinges on integrating diverse datasets and local geological considerations. In Poland, current assessments face significant challenges rooted in the lack of availability and comprehensiveness of data, which are compounded by limited access to the research sector and collaborative industry-academia R\&D initiatives. Due to the heterogeneous nature of available geological data, varying assessment methodologies across different studies, and the limited number of comparable site-specific evaluations, formal statistical meta-analysis was not applied in this review. Consequently, basin-scale assessments and site-specific studies must account for confidence intervals and reliability caveats until new seismic surveys, well-site measurements, and core data become available. We emphasize to stakeholders that the current state of knowledge and their tentative high-level storage capacity estimates, such as those presented in Table \ref{table:storagelist}, are still insufficient for making definitive business decisions, necessitating further scrutiny and validation.

\paragraph{Geological Data:} Subsurface characterization relies on core samples, geophysical well logs, and seismic surveys. In Poland, it is often sourced from the Polish Geological Institute (PGI-NRI), Polskie Górnictwo Naftowe i Gazownictwo (PGNiG), and Grupa LOTOS. These datasets, to their extent, provide input into reservoir properties across Poland’s sedimentary basins. Legacy Soviet-era 2D seismic data (1970s–1980s), reprocessed with modern algorithms, supplements contemporary 3D surveys. However, its coarse resolution limits detailed modeling of structural traps and reservoir heterogeneity. Recent efforts to digitize and reinterpret this data have improved coverage, but gaps persist in the majority of regions, particularly regarding the characterization of deep saline aquifers.

\paragraph{Hydrogeological Data:} The Poland Hydrogeological Atlas \cite{gorecki2015atlases,joywiakmapy} provides preliminary salinity data (typically >10,000 mg/L for viable aquifers) and pressure gradients, essential for assessing CO\textsubscript{2} solubility, plume migration, and injectivity. Over-pressured formations are excluded in many relevant studies, which is consistent with global practices aimed at mitigating injection challenges and ensuring operational feasibility when accurate information is absent.

\paragraph{Thermodynamic Data:} Geothermal gradients, often ranging from 25–35°C/km, dictate CO\textsubscript{2} phase behavior (supercritical above 31°C and 73.8 bar) and influence storage efficiency at varying depths. These data, combined with pressure-temperature profiles from well logs (where available), enable computation of CO\textsubscript{2} density and viscosity under reservoir conditions, aligning with international standards \cite{goodman2011us}.

Reprocessed Soviet-era 2D seismic data, despite modern enhancements, falls short of the resolution required for detailed subsurface mapping and dynamic simulations. In the Polish Lowland Basin (PLB), over 1,200 legacy wells, many of which were drilled before the 1980s, exhibit uncertain cement integrity. This necessitates the use of cement evaluation logs and pressure tests to quantify leakage risks \cite{Tarkowski2009, celia2015status}.

As detailed in subsequent sections, many study areas (e.g., Region V in Podlasie, Figure \ref{fig:F7_basins}) face geological and geophysical data scarcity, including limited 3D seismic coverage and reliance on 2D surveys. The Baltic Sea analogs suggest a speculative theoretical capacity. Similarly, Study Area VIII’s 1.64 Gt estimate, derived from Swedish Cambrian reservoirs, is limited by sparse well penetrations beyond the B3 oil field \cite{wojcicki2014assessment}. These extrapolations exhibit wide confidence intervals (P10--P90 ranges of $\pm$30--50\%) due to insufficient subsurface validation. Current basin-scale and prospect-specific studies must thus incorporate probabilistic uncertainty analyses to define reliability bounds.

\section{Geological Setting and Sedimentary Basins in Poland}

\begin{figure}
 	\centering
 	\includegraphics[width=1.0\textwidth]{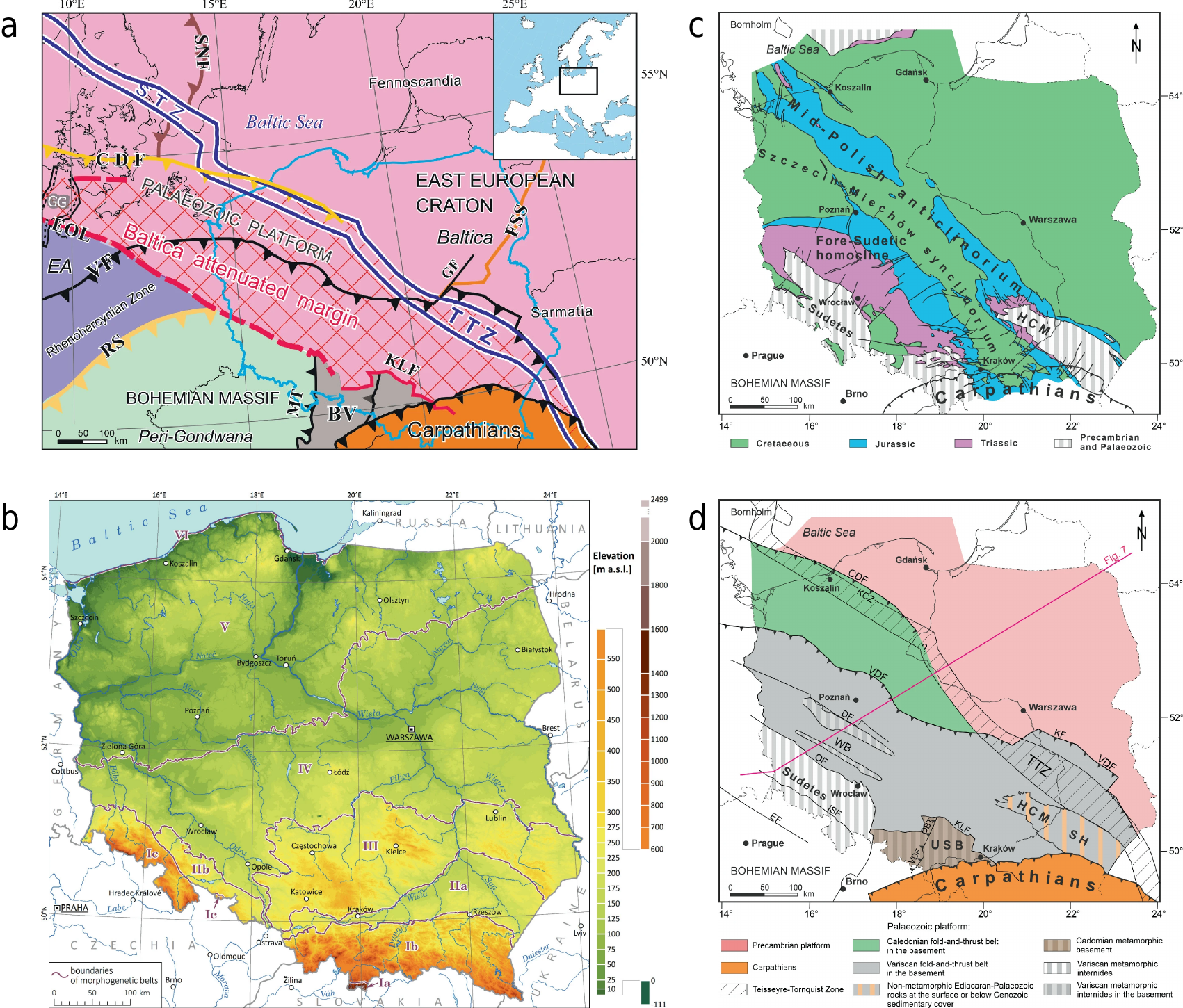}
    \caption{Geological and geomorphological overview of Poland and its tectonic framework.\\ 
    (a) Crustal domains at the transition from the Precambrian East European Platform to the Palaeozoic Western European Platform and Western Carpathians, with major structural elements and the Polish boundary (blue outline). The inset map shows the location within Europe. Abbreviations: BV—Brunovistulicum; CDF—Caledonian Deformation Front; EA—East Avalonia; EOL—Elbe–Odra Lineament; FSS—Fennoscandia-Sarmatia Suture; GF—Grójec Fault; GG—Glückstadt Graben; KLF—Kraków–Lubliniec Fault; MT—Moldanubian Thrust; RS—Rheic Suture; SNF—Sveconorwegian Front; STZ—Sorgenfrei–Tornquist Zone; TTZ—Teisseyre–Tornquist Zone; VDF—Variscan Deformation Front (modified from \cite{Mazur2024, mazur2021pre}). 
    (b) Altitudinal diversity of Poland with morphogenetic belts. Mountains: Ia—High Mountains (Tatra Mts.); Ib—Medium–High Mountains of the Alpine System (Beskidy Mts.); Ic—Medium–High Mountains of the Variscan System (Sudetes). Fore-mountain areas: IIa—Sub-Carpathian Basins; IIb—Sudetic Foreland. III—Uplands; IV—Old Glacial Relief (Central Poland); V—Young Glacial Relief (Northern Poland); VI—Coastal Zone (modified from \cite{Migoń2024}). 
    (c) Pre-Cenozoic geology of Poland, highlighting the Permian–Mesozoic Polish Basin structure during Late Cretaceous inversion. HCM—Holy Cross Mountains (modified from \cite{dadlez2000geological,
    Mazur2024}).  
    (d) Tectonic provinces of Poland, excluding Permian–Cenozoic cover. Abbreviations: CDF—Caledonian Deformation Front; DF—Dolsk Fault; EF—Elbe Fault; HCM—Holy Cross Mountains; ISF—Intra-Sudetic Fault; KCZ—Koszalin–Chojnice Zone; KF—Kock Fault; KLF—Kraków–Lubliniec Fault; OBT—Orlová–Boguszowice Thrust; OF—Odra Fault; SH—San Horst; TTZ—Teisseyre–Tornquist Zone; USB—Upper Silesia Block; VDF—Variscan Deformation Front; WB—Wolsztyn Block \cite{aleksandrowski2017nowych, Mazur2024}).} 
 \label{fig:F6_tectonic2}
 \end{figure}

Poland's subsurface is a complex geological mosaic shaped by a dynamic tectonic history and diverse stratigraphy. Formed through Phanerozoic sedimentation, it bears the imprint of the Caledonian, Variscan, and Alpine orogenic cycles, followed by post-orogenic subsidence and glacial overprinting \cite{oliver1993u,mazur2006variscan,kutek1972holy}. These events created a variety of sedimentary basins, differing in age, lithology, structure, and petrophysical properties \cite{krzywiec2001contrasting}. Figure~\ref{fig:F6_tectonic2} illustrates Poland's geological and geomorphological overview and its tectonic framework.

\subsection{Tectonic Evolution}

\subsubsection*{Caledonian Orogeny}

The Caledonian progeny (Fig.~\ref{fig:F6_tectonic2}a) , an early Paleozoic tectonic event, was critical in forming Poland's foundational geological structures. During this period, the closure of the Tornquist Sea led to dramatic crustal deformation, subsidence, and intense sediment accumulation. This orogeny laid the foundation for complex stratigraphic layers, resulting in thick sequences of marine and terrestrial deposits interspersed with volcanic materials. These formations, now buried at significant depths, offer potential reservoirs for carbon storage, although their heterogeneity requires rigorous characterization \cite{znosko1986polish,johnston1994evidence}.

\subsubsection*{Variscan Orogeny}

The late Paleozoic era was marked by the Variscan progeny (Fig.~\ref{fig:F6_tectonic2}a), characterized by intense mountain building and tectonic activity. This era was affected by widespread folding, faulting, and the emplacement of extensive sedimentary sequences, especially pronounced in the southwestern regions of Poland. The Variscan orogenic belt facilitated the development of thick coal-bearing strata and significant clastic deposits, which not only have historical economic significance due to coal mining but also present potential for enhanced gas recovery and CO\textsubscript{2} sequestration. These structures, characterized by complex folding and thrusting, require advanced geophysical techniques for precise assessment \cite{mazur2006variscan}.

\subsubsection*{Alpine Orogeny}

The Mesozoic to Cenozoic alpine orogeny further redefined the tectonic architecture of the region, especially influencing the Carpathian domain. This phase was associated with the evolution of foreland basins, such as the Carpathian Foredeep, which is notable for its significant structural complexity and is an ideal feature for geological storage. The intense tectonic reworking during the Alpine orogeny resulted in varied lithologies and created multiple stratigraphic traps, enhancing the potential for secure CO\textsubscript{2} storage. The resulting ultramafic sequences and flysch deposits, coupled with synorogenic turbidites, offer both opportunities and challenges in the realm of carbon capture and storage strategies \cite{kutek1972holy}.

\subsubsection*{Post-Orogenic Subsidence and Glacial Impact}

Following these tectonic upheavals, periods of post-orogenic subsidence contributed to the development of extensive sedimentary basins, such as the Polish Basin. Glacial activity during the Quaternary period had a significant impact on erosion patterns and deposition processes, shaping Poland's current subsurface characteristics. Glacial deposits and moraines significantly contribute to the modern landscape, influencing porosity and permeability patterns that are crucial for evaluating the efficacy of storage sites. Understanding these glacial imprints, along with the transgressions and regressions that influence sediment distribution and compaction \cite{marks2005pleistocene,marks2012timing}.

\subsection{Tectonic setting}

Poland's subsurface is shaped by three major tectonic domains: the East European Craton, the Teisseyre-Tornquist Zone, and the Carpathian orogen (Fig.~\ref{fig:F6_tectonic2}). The convergence of these domains creates a complex geological landscape, presenting both opportunities and challenges for carbon sequestration. The diverse subsurface characteristics across these regions require thorough geological assessments to realize their carbon storage potential, a critical component of strategies to mitigate climate change \cite{oszczypko2006carpathian}.

The lithostratigraphic architecture is primarily defined by these tectonic domains (Figs. \ref{fig:F6_tectonic2}–\ref{fig:F7_basins}).

\begin{itemize}
    \item \textbf{East European Craton (EEC):} This stable Precambrian basement underpins the northeastern part of Poland, including the Baltic Basin and the Polish Lowlands. EEC stability has helped preserve extensive sedimentary sequences spanning from the Cambrian to the Quaternary periods. These sequences predominantly consist of sandstones, shales, and carbonates, which have undergone minimal tectonic deformation, making them favorable for storage candidates \cite{kusek2021geological}.

    \item \textbf{Trans-European Suture Zone (TESZ):} Serving as a prominent NW-SE trending boundary, the TESZ delineates the transition between the ancient EEC and the younger Paleozoic platforms to the southwest. This zone is characterized by a complex assemblage of terranes and suture zones resulting from past collisional events. The TESZ has significantly influenced sediment dispersal patterns and basin evolution, acting as a conduit for sediment transport and affecting subsidence rates \cite{kusek2021geological}.

    \item \textbf{Carpathian Orogen:} Located in southern Poland, the Carpathian Orogen is a dynamic Cenozoic fold-thrust belt formed due to the Alpine orogenic processes. This region encompasses the Carpathian Foredeep, a flexural foreland basin characterized by intricate structural and stratigraphic patterns. The evolution of the orogen has been pivotal in shaping the sedimentary environments and the resource potential of the area \cite{miu2021assessment}.
\end{itemize}

These tectonic domains have nurtured the formation of several crucial sedimentary basins in Poland that are of interest for CO\textsubscript{2} storage (see also Fig. \ref{fig:F7_basins}), including:
\begin{itemize}
    \item Polish Lowlands Basin (PLB): A Permian-Mesozoic intracratonic basin.
    \item Carpathian Foredeep (CF): A Neogene foreland basin.
    \item Baltic Basin (BB): An offshore passive margin setting with Cambrian-Ordovician successions.
    \item Fore-Sudetic Monocline (FSM): A Permian rift-related depocenter.
\end{itemize}

\subsection{Major Sedimentary Basins}

Poland's four major sedimentary basins, shaped by the interplay of tectonic activity and sedimentary processes, are defined by distinct geodynamic histories and geological characteristics (Fig. \ref{fig:F7_basins}). These basins exhibit unique sedimentary sequences, offering significant potential for geoenergy exploitation and carbon sequestration. Their complex geological features, formed through tectonic upheaval and sediment deposition, necessitate tailored exploration programs to unlock their full resource potential. Such assessments are crucial for advancing sustainable energy strategies and supporting Poland's efforts to mitigate climate change through effective carbon management.

\subsubsection{Polish Lowlands Basin (PLB)}

The PLB is a significant Permian-Mesozoic intracratonic basin that extends across much of north-central Poland. Characterized by a voluminous succession of sedimentary rocks, including vast deposits of sandstones, shales, and carbonates, the PLB reflects a dynamic sedimentary environment facilitated by major tectonic events and subsequent subsidence \cite{tarkowski2008co2,sowizdzal2022future}. The evolution of this basin has undergone phases of extensional tectonics, followed by periods of thermal subsidence, resulting in the creation of substantial structural traps that are favorable for hydrocarbon entrapment. Among the notable hydrocarbon-rich provinces within the PLB, we identify:

\begin{itemize}
    \item \textbf{Pomerania Province:} Situated in the northwest, Pomerania is notable for its potential hydrocarbon reserves, highlighted by extensive stratigraphic and basin analyses. Preliminary assessments indicate promising exploration targets, particularly within the Triassic and Jurassic sequences.

    \item \textbf{Wielkopolska Province:} Located in west-central Poland, this province showcases several hydrocarbon-bearing formations. Ongoing exploration actions focus on deep geological units, including Permian and Carboniferous strata, with the aim of discovering new viable oil and gas reserves within identified structural trends.

    \item \textbf{Małopolska Province:} Spanning southern Poland, Małopolska incorporates both conventional and unconventional hydrocarbon resources. Its stratigraphy reveals a promising potential given its diversified depositional settings, which encompass Paleozoic to Mesozoic sequences suitable for continuous exploration.
\end{itemize}

\subsubsection{Carpathian Foredeep (CF)}

The Carpathian Foredeep is a Neogene foreland basin located along the northern fringe of the Carpathian Mountains. It epitomizes a complex stratigraphic arrangement anchored by Miocene clastic deposits, predominantly sandstones and shales, which were laid down under foreland basin conditions amidst an active compressional regime \cite{chmielowska2021prospects}. The intricate geological setup fosters a variety of structural traps, including those associated with thrust faults and fold belts. The basin's hydrocarbon potentials are primarily concentrated in the following areas:

\begin{itemize}
    \item \textbf{Lublin Province:} Situated in the east, Lublin has attracted significant exploration interest. Its Miocene strata, rich in potential reservoirs, have already revealed several hydrocarbon fields beneath structural traps, where lead activities target both conventional and unconventional reserves.

    \item \textbf{Gdańsk Province:} Located in northern Poland, Gdańsk is distinguished by the hydrocarbon potential within the Miocene and pre-Miocene formations. Ongoing evaluations seek to quantify recoverable resources through integrating geophysical surveys and drilling techniques.
\end{itemize}

\subsubsection{Baltic Basin (BB)}

The Baltic Basin is an offshore passive margin setting that extends beneath the Southern Baltic Sea, encompassing a remarkable Cambrian-Ordovician succession. Significant early Paleozoic sedimentation has produced a classical sequence of platform carbonates and siliciclastics, whose structural integrity affords promising sites for hydrocarbon exploration. Enhanced by regional geophysical data, the potential of the basin is encapsulated in its deep sedimentary sequences, which indicate dependable reservoirs \cite{malik2024assessing}.

\subsubsection{Fore-Sudetic Monocline (FSM)}

The Fore-Sudetic Monocline (FSM) is a Permian rift-related depocenter in southwestern Poland. This structural feature was formed during the extensional tectonics associated with the Permian rifting phase, resulting in the deposition of thick sequences of volcaniclastic and sedimentary rocks. The FSM is characterized by its structural simplicity but has complex internal stratigraphy that offers considerable prospects for geological exploration, particularly within basal conglomerates and sealed structural highs \cite{papiernik2015structural}.

\begin{figure}[h]
    \centering
   \includegraphics[width=1\textwidth]{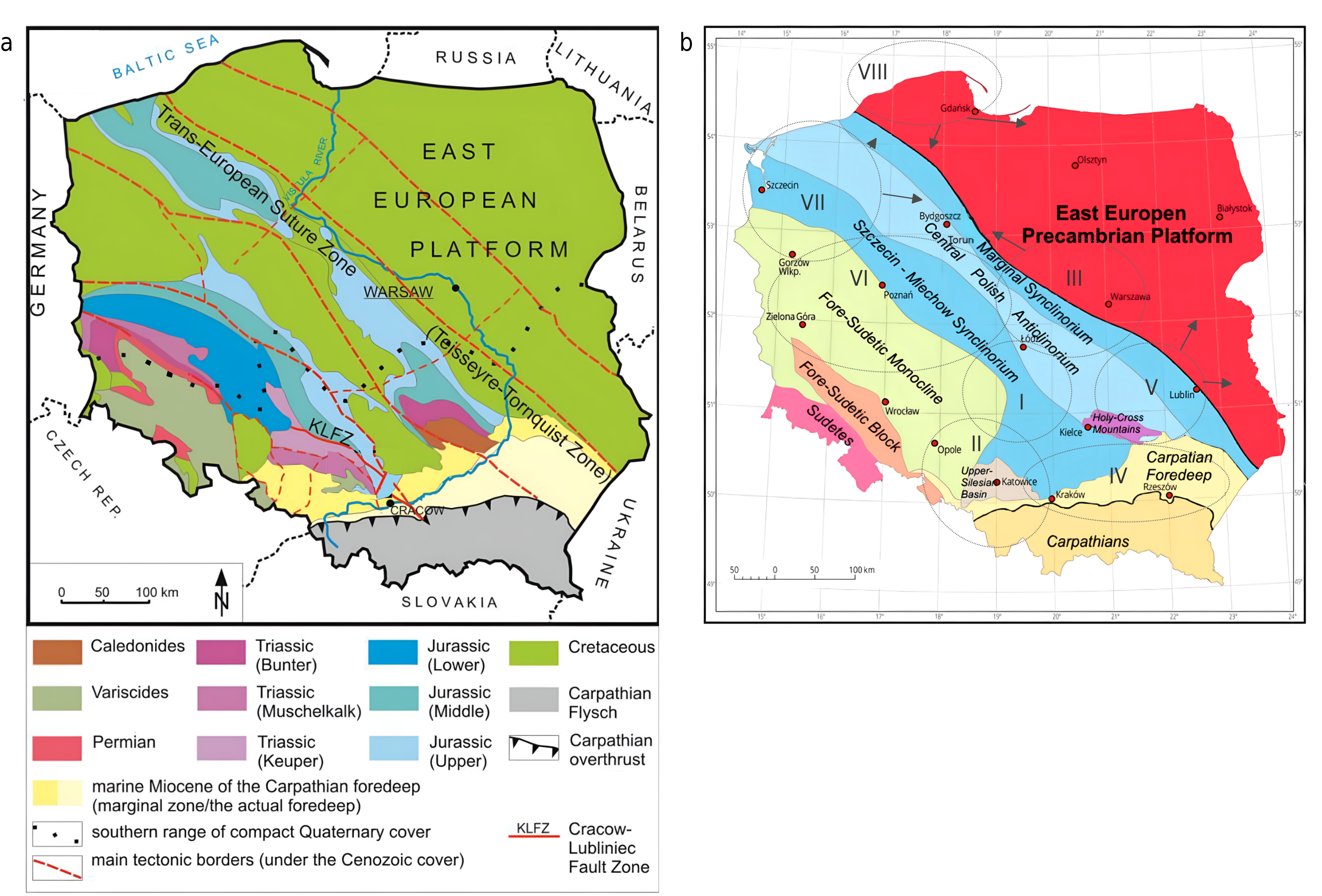}
    \caption{Geological framework and subsurface sedimentary basins in Poland.\\ (a) Simplified subsurface geological map of Poland, excluding Cenozoic cover, highlighting Mesozoic and Paleozoic formations relevant to CO\textsubscript{2} storage assessments (modified from \cite{mikulski2020trace,gruszczyk1990zinc}). (b) Distribution of geological basins and evaluated storage regions, showing eight primary saline aquifer regions (I--VIII, following \cite{wojcicki2014assessment}). Storage targets include: Permian--Mesozoic formations in four regions (Bełchatów, Warsaw--Mazovia, Greater Poland--Kujawy, NW Poland); Paleozoic formations in five regions (Upper Silesian Coal Basin, Lublin--Podlasie, Łeba Elevation, Baltic Sea economic zone, NE Poland); and mixed Mesozoic--Paleozoic formations in the Carpathian Overthrust and Foredeep. Additional storage opportunities exist in depleted hydrocarbon fields (western and southeastern Poland) and unmineable coal seams (Upper Silesian Coal Basin), though the latter require further technical evaluation.}
    \label{fig:F7_basins}
\end{figure}

\section{Storage Potential of Sedimentary Basins in Poland}

Poland’s sedimentary basins offer diverse geological settings suitable for CO\textsubscript{2} storage. This section synthesizes stratigraphic, structural, and petrophysical data and interpretation sourced primarily from Poland’s National CCS Assessment Project \cite{wojcicki2014assessment}, evaluating storage potential across eight key study areas (I--VIII), as shown in Figure \ref{fig:F7_basins}. A summary is provided in Table \ref{table:storagelist}, recently updated through the EU CCUS ZEN project \cite{Ringstad2023CCUSZEN}. These assessments are primarily at the theoretical capacity level and correspond to SRL 1--3. These regions, however, may face potential land-use restrictions due to regulatory zoning, environmental considerations, Natura 2000 provisions, proximity to population centers, safety and security risks, and conflicts with other subsurface energy applications, further constraining these high-level theoretical CO\textsubscript{2} storage estimates presented below.

\subsection{Onshore Basins}
\subsubsection{Polish Lowlands Basin (PLB) – Study Areas I, III, VI, VII}

The Permian-Mesozoic PLB, covering 72\% of Poland’s territory, accounts for approximately 85\% of the country’s theoretical CO\textsubscript{2} storage potential. Key sub-regions include:

\begin{itemize}
    \item \textbf{Bełchatów Zone (Study Area I):} Characterized by Lower Jurassic anticlinal traps such as Budziszewice-Zaosie and Wojszyce, these zones exhibit porosity ranging from 15–25\% and are positioned in supercritical CO\textsubscript{2} conditions (35–70°C at depths of 775 to 2,265 m). A 100-m-thick Toarcian claystone with nano-darcy-scale permeability ensures caprock integrity. \textbf{Key risk}: The proximity to Łódź potable aquifers necessitates targeted geochemical monitoring of brine-freshwater interfaces.

    \item \textbf{Mazovia Trough (Study Area III):} Featuring a multi-reservoir Jurassic-Cretaceous system (Bielsk-Bodzanów, Sierpc anticlines) with net sand thicknesses of 20-30 m, dynamic simulations indicate a capacity to handle industrial emissions from Warsaw-Płock, totaling 2.6 Gt. \textbf{Conflict}: This area overlaps with shale gas exploration licenses on the Lublin Basin periphery.

    \item \textbf{Fore-Sudetic Monocline (Study Area VI):} The Poznań Trough Megastructure houses Rotliegend volcaniclastics with a capacity of 634 Mt. Zechstein caprock integrity needs reinforcement, as 412 legacy wells exist that likely require re-cementation. \textbf{Innovation}: There is potential for CO\textsubscript{2}-EGR in the adjacent depleted Wilków gas field with a 13.6 Mt capacity.

    \item \textbf{NW Poland (Study Area VII):} The Choszczno-Suliszewo Triassic-Jurassic structure boasts a significant theoretical capacity. A standout feature is its high permeability, enabling million-ton per annum injection rates; however, approximately 40\% of this area intersects with NATURA 2000 zones, imposing limitations on surface infrastructure development.
\end{itemize}

\subsubsection{Carpathian Foredeep (CF) – Study Area IV}

The Carpathian Foredeep, a thrust-and-fold belt, offers dual storage mechanisms in:

\begin{itemize}
    \item \textbf{Badenian Sandstones:} Deltaic turbidites with porosities of 18–22\% are prominent at the Skoczów-Czechowice site (Study Area II). Seal integrity might be initially verified in the Menilite shales, although fault reactivation necessitates de-risking and detailed site characterization.

    \item \textbf{Carboniferous-Devonian Carbonates:} The fractured Grobla and Niepołomice structures offer further major pore volume capacity, with enhanced permeability of the karst (50–200 mD) that facilitates the capture of dissolution

\end{itemize}

\subsubsection{Lublin-Podlasie Region (Study Area V)}

This region houses underexplored aquifers with a high-level theoretical static evaluation of several hundred megatons:

\begin{itemize}
    \item \textbf{Carboniferous C3 Sands:} With a thickness of 30 meters and porosity ranging from 5–15\%, the Stężyca IG-1 well shows promising reservoir properties. However, limited seismic coverage introduces uncertainty.
    
    \item \textbf{Cambrian Reservoirs:} The Podlasie regional aquifer, spanning 650 km\textsuperscript{2} with a 200 m net sand thickness, could exhibit low storage efficiency in preliminary assessment due to structural complexity, necessitating detailed seismic surveys to characterize reservoir units, structures, and enable dynamic capacity assessments.
\end{itemize}

\subsection{Offshore Basins – Study Area VIII}

\subsubsection{Baltic Basin}

The Cambrian Deimena Formation sands within Poland’s exclusive economic zone offer cross-border potential:

\begin{itemize}
    \item {Block B (Offshore):} With salt-cored anticlines in the B3 field, this area could offer huge practical storage capacity. A high injectivity is expected within these units.

   \item {Block E (Onshore):} The Elbląg region’s Cambrian sands, with significant storage capacity, align with the D6 structure in Kaliningrad, necessitating EU-mediated storage agreements and resolution of cooperation conflicts outlined in the introduction.

   \item The Middle Cambrian Debkowska Formation offers cross-border CO\textsubscript{2} storage potential, extending from Polish sandstone units into the Swedish aquifers of the Dalders Monocline (Faludden Formation). Using a conservative 2\% storage efficiency for a semi-open aquifer, the recent CCUS ZEN evaluations estimate 747 Mt of storage capacity on the Swedish side and 188 Mt on the Polish side \cite{Gravaud2023}. Additionally, two depleted hydrocarbon fields with structural traps in the same Polish unit provide further storage potential.
\end{itemize}

\begin{figure}[h!]
    \centering
    \includegraphics[width=0.9\textwidth]{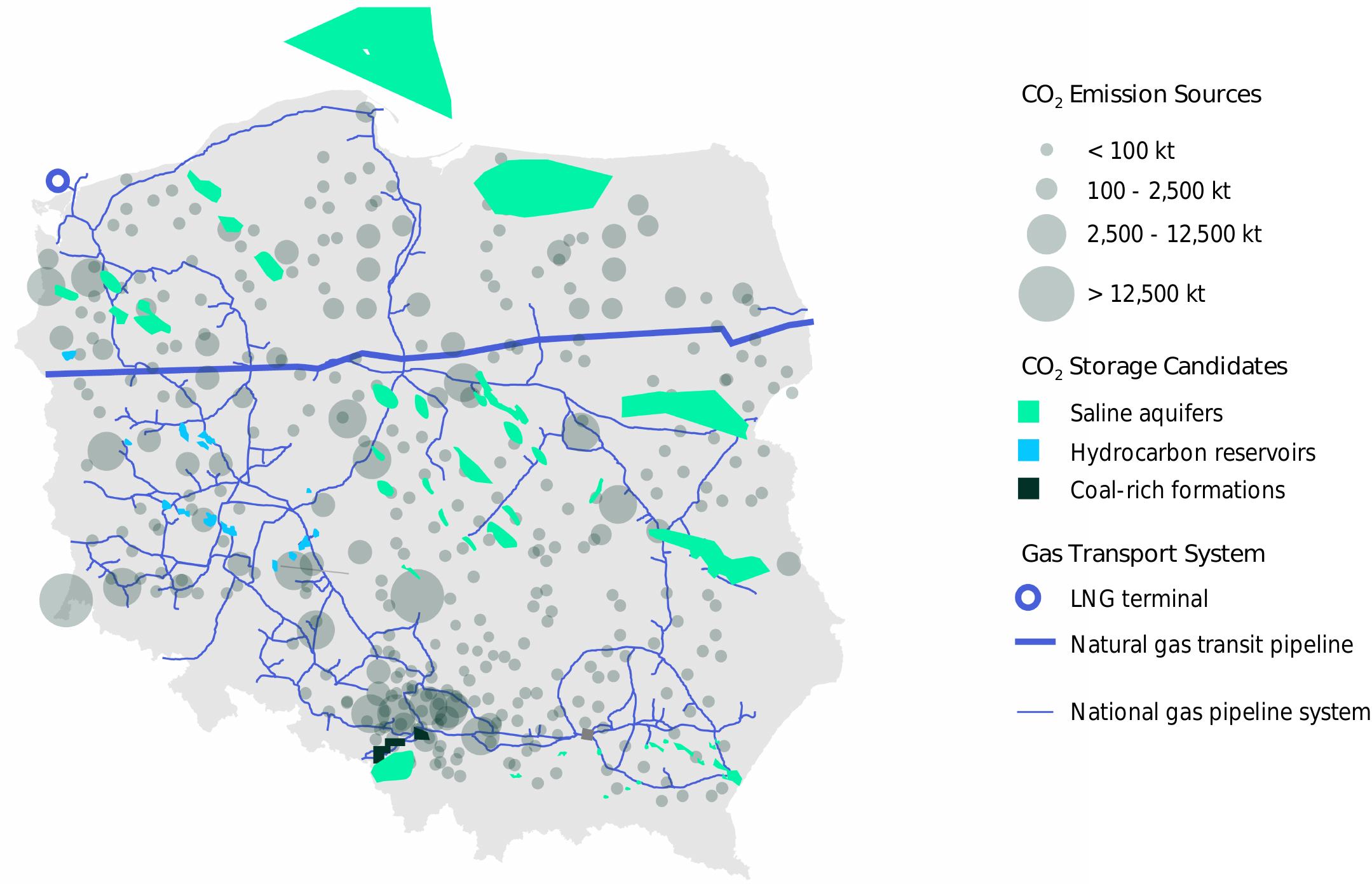}
    \caption{Spatial distribution of major CO\textsubscript{2} emission sources and potential geological storage sites in Poland.\\ The map shows 102 industrial facilities emitting >100 kt CO\textsubscript{2}/year (totaling 189,183 kt/year in 2023), categorized by sector: power generation dominates with 83 facilities (143,272 kt/year, ~80\% of total emissions), followed by cement (9 facilities, 11,723 kt/year), chemicals (15 facilities, 9,289 kt/year), and other industries including refineries and steel production. Potential storage targets include deep saline aquifers, depleted hydrocarbon fields, and unminable coal seams. Poland's Natura 2000 protected areas (~20\% of national territory) may constrain surface infrastructure development in ecologically sensitive regions. The spatial analysis illustrates proximity relationships between emission sources and sequestration opportunities for CCS deployment planning. Adapted from Polish Geological Institute's CO\textsubscript{2} Storage Atlas and \cite{wojcicki2014assessment,nooraiepour2025norwegian,Ringstad2023CCUSZEN,IEA_Poland_Emissions_2023,McKinsey2020}.}
      \label{fig:F8_storage}
\end{figure}

\subsection{Geological Carbon Storage Potential}

Figure \ref{fig:F8_storage} and Table \ref{table:storagelist} provide a high-level overview of potential geological storage candidates with SRL 1--3 in Poland. Poland’s sedimentary basins, including the Polish Lowlands Basin, Carpathian Foredeep, and Fore-Sudetic Monocline (Figure \ref{fig:F7_basins}), indicate a significant but highly uncertain combined theoretical storage capacity in saline aquifers of approximately 10~Gt ($\pm$25\%, 7.5--12.5~Gt).

These aquifers exhibit favorable geological properties tailored for CO\textsubscript{2} sequestration. The Polish Lowlands Basin, with its Rotliegend and Buntsandstein sandstones, may theoretically contribute 6.7 GtCO\textsubscript{2}, characterized by porosities of 12--18\% and a broad permeability range at deep enough depths of scCO\textsubscript{2} sequestration. The Carpathian Foredeep offers 3.0--5.0 GtCO\textsubscript{2} within Badenian Sandstones and Sarmatian Carbonates. The smaller Fore-Sudetic Monocline adds 0.5--1.0 GtCO\textsubscript{2} from Rotliegend Volcanics and Zechstein Limestones, despite lower petrophysical properties. Onshore storage in PLB and CF, as shown in Figs. \ref{fig:F7_basins}–\ref{fig:F8_storage}, benefits from proximity to industrial centers, reducing transportation costs and enabling integration with existing infrastructure. The substantial storage capacities of the Rotliegend and Jurassic formations are particularly appealing for large emitters. 

Depleted hydrocarbon reservoirs in these onshore basins (Figure \ref{fig:F8_storage}) supplement this capacity with practical CO\textsubscript{2} storage estimates of 0.1--0.5~Gt, leveraging existing infrastructure and geological data from prior hydrocarbon extraction. The PLB contributes the majority, with the CF providing additional capacity. Offshore, the saline aquifers of the Baltic Basin offer significant theoretical capacity, as noted earlier. These reservoirs, composed of Cambrian-Ordovician sandstones, benefit from high porosities (20--28\%) and high permeability values at comparatively shallow depths. Recent evaluations by an EU project consortium \cite{Ringstad2023CCUSZEN, Gravaud2023} identified 55 aquifer units and 5 hydrocarbon fields, with a storage readiness level of 2--3, yielding a total CO\textsubscript{2} storage capacity of 8.9 Mt at a 20\% storage efficiency.

Coal seams in the Upper Silesian Coal Basin (Figs. \ref{fig:F7_basins}-\ref{fig:F8_storage}) might offer potential CO\textsubscript{2} storage. Although individual sites have limited capacity, developing multiple sites could cumulatively contribute to increased storage capacity, further incentivized by the economics of methane recovery.

Poland’s high-level CO\textsubscript{2} storage capacity estimates range from approximately 8.9 Gt (SRL 2--3) to 10.2--12.7 Gt (SRL 1--2), provisionally up to 15 Gt, though technical and regulatory challenges, discussed in subsequent sections, persist. A hybrid approach integrating onshore and offshore storage, supported by rigorous monitoring, advanced reservoir simulation, and adaptive management, is essential to optimize sequestration and achieve decarbonization goals.

\section{Potential for Carbon Mineralization}

Poland's geological framework, spanning the East European Craton, Bohemian Massif, and Carpathian orogenic belt, offers a diverse substrate for carbon mineralization in mafic and ultramafic rocks. These rocks, rich in magnesium-, calcium-, and iron-bearing silicates (e.g., olivine, pyroxene, serpentine), react with CO\textsubscript{2} to form stable carbonate minerals like magnesite and calcite, enabling permanent sequestration through mineral trapping. Although Poland lacks extensive basalt plateaus or ophiolite complexes, mafic and ultramafic formations in the Sudetes Mountains, Upper Silesia, Holy Cross Mountains, Gorzów Wielkopolski region, and Tertiary volcanic provinces may show theoretical potential (SRL 1) for carbon mineralization. Due to limited site-specific geochemical data, this assessment relies on Carbfix thermodynamic databases \cite{voigt2018evaluation}. Table~\ref{tab:carbon_mineralization_summary} summarizes the preliminary data, outlining the carbon mineralization potential and constraints for distributed emitters.

\subsection{Regional Distribution and Tectonic Context}
The Sudetes Mountains in southwest Poland host the most significant mafic and ultramafic exposures, notably the Ślęża Ophiolite near Sobótka and the Nowa Ruda ophiolite, both part of the Variscan oceanic lithosphere (ca. 353 $\pm$ 21 Ma, Sm-Nd dating). The Ślęża Ophiolite, spanning approximately 20 km\textsuperscript{2}, features a pseudostratigraphy transitioning from serpentinites and gabbros in the south to metabasite lavas and pillow basalts in the north, with recent discoveries of epidosites (quartz + epidote + titanite) in its sheeted dyke complex indicating a supra-subduction zone affinity. Nowa Ruda, although smaller (<10 km²) and tectonically fragmented, mirrors this composition with serpentinites and gabbros, which are also part of the Variscan suture.

In central Poland, the Holy Cross Mountains, part of the Trans-European Suture Zone, are characterized by Devonian to Carboniferous mafic volcanic rocks (approximately 400–300 Ma), including basalts and diabases, which are linked to Variscan rifting and extensional tectonics. Northwest Poland’s Gorzów Wielkopolski Block, within the Fore-Sudetic Monocline, contains Permian Rotliegend metavolcanic rocks (andesite-basalts and andesites, 285 $\pm$ 5 Ma, K-Ar dating), overlying Paleozoic basement and underlying Zechstein evaporites, providing a deep-seated target for mineralization. The Tertiary volcanic provinces in the Kaczawa Mountains and Złotoryja region feature Neogene basaltic lava flows, pyroclastic deposits, and volcanic necks, such as Ostrzyca, surrounded by Permian sandstones and conglomerates, with pervasive cooling fractures that enhance fluid migration. These formations are confined to tectonically active or ancient orogenic zones.

\subsection{Depth, Distribution, and Petrophysical Properties}
Depth and distribution vary depending on the tectonic context. The Ślęża Ophiolite ranges from surface exposures to depths of several kilometers, with some sections buried under Mesozoic-Cenozoic sediments at 1–5 km, as inferred from Soviet seismic profiles. Nowa Ruda follows a similar pattern, while Holy Cross Mountain exposures are shallow (<100 m), accessible for surface-based studies. Gorzów Wielkopolski’s Permian metavolcanics, sampled from boreholes (e.g., Namyoelin 1, Witnica 1, Dzieduszyce 1), lie beneath thick sedimentary cover, reaching depths of 1–3 km, overlying Lower Carboniferous clastics and sealed by evaporites. Kaczawa and Złotoryja basalts, typically <200 m thick, are near-surface or shallowly buried, with Ostrzyca’s fracture-rich structure enhancing accessibility. Laterally, these rocks are scattered: Ślęża at 20 km\textsuperscript{2}, Nowa Ruda and Upper Silesian basalts at <10 km\textsuperscript{2} each, and Gorzów Wielkopolski’s metavolcanics forming patchy lenses. Vertical heterogeneity, including faults and variable lithological continuity, necessitates high-resolution mapping to optimize CO\textsubscript{2} injection depths and mitigate risks like induced seismicity.

Petrophysical properties, age, and composition govern CO\textsubscript{2} reactivity. Generally, mafic rocks (basalts, gabbros) exhibit densities of 2.7–3.0 g/cm\textsuperscript{3}, porosities of 1–10\% (up to 15\% in fractured zones), and permeabilities of 10\textsuperscript{–19} to 10\textsuperscript{–15} m\textsuperscript{2} (10–100 mD in fractures), driven by plagioclase, pyroxene, and minor olivine (45–52\% SiO\textsubscript{2}). Based on global studies, ultramafic rocks (peridotites, serpentinites) are denser (3.0–3.3 g/cm\textsuperscript{3}), with porosities of 5–15\% and permeabilities of 10\textsuperscript{–16} to 10\textsuperscript{–13} m\textsuperscript{2} (<1 mD intact), rich in olivine (MgO >18–50 wt\%), pyroxene, and serpentine (SiO\textsubscript{2} <45\%), plus accessory chromite and magnetite. Ślęża’s chromitites show variable Cr\# (0.41–0.68) and Mg\# (0.62–0.83), with ferrogabbros containing up to 14.6 vol\% magnetite-ilmenite. Gorzów Wielkopolski’s andesite-basalts preserve subalkaline basalt signatures despite low-grade metamorphism (<200\textdegree C, 2 kbar), with assemblages like pumpellyite and laumontite. Kaczawa basalts, fine-grained and fractured, may have enhanced carbonation kinetics. 

\subsection{Age, Composition, and Alteration}
Ages span Neoproterozoic to Early Paleozoic for Sudetes, Devonian-Carboniferous for Upper Silesia and Holy Cross, Permian for Gorzów Wielkopolski, and Neogene for Kaczawa/Złotoryja, reflecting Poland’s complex tectonic evolution. The age relationships are essential to establish reliably via targeted investigations for understanding alteration history and current mineral stability, which directly affects CO\textsubscript{2} reactivity and storage capacity in these formations. Compositions include mafic rocks dominated by plagioclase, pyroxene, and minor olivine (SiO\textsubscript{2} 45–52 wt\%), and ultramafic rocks rich in olivine (MgO 18–50 wt\%), pyroxene, and serpentine (SiO\textsubscript{2} <45 wt\%), with accessories like chromite and magnetite. Alteration and weathering significantly influence suitability. Serpentinization dominates Sudetic ultramafic rocks, converting olivine to serpentine, magnetite, and brucite, increasing porosity but depleting reactive silicates, with natural magnesite and calcite veins forming under Poland’s temperate climate. Gorzów Wielkopolski’s metavolcanics show pervasive low-grade metamorphism, replacing primary minerals with corrensite and zeolites, leaving rare clinopyroxene and Cr-spinel. Mafic rocks show milder alteration (chloritization, epidotization), with weathering occurring shallowly (<100 m) and being fracture-enhanced in the Kaczawa basalts, although it is slow compared to tropical rates.

\subsection{Suitability for CO\textsubscript{2} Mineralization}

CO\textsubscript{2} mineralization suitability hinges on reactivity, volume, and accessibility. As established in experimental studies, ultramafic rocks offer high uptake (0.62 t CO\textsubscript{2}/t olivine, 0.4–0.5 t CO\textsubscript{2}/t serpentine), with fresh peridotites rapidly forming magnesite, while mafic rocks provide 0.3–0.4 t CO\textsubscript{2}/t basalt via Ca/Mg-rich minerals. Preliminary capacity estimates, based on conservative rock volumes and efficiency factors ranging from 1\% to 20\%, suggest Ślęża’s 20 km\textsuperscript{2} might theoretically and provisionally store 1–10 Mt CO\textsubscript{2}, Nowa Ruda and Holy Cross sites 5–20 Mt each, Gorzów Wielkopolski and Kaczawa/Złotoryja similarly limited. Low ultramafic permeability necessitates hydraulic fracturing, while fractured mafic rocks (e.g., Ostrzyca) enhance injectivity, though deep reservoirs (>1 km) elevate drilling costs. In-situ mineralization suits both rock types, with supercritical CO\textsubscript{2} viable for basalts as well. 

Poland’s mafic and ultramafic rocks may theoretically provide localized CO\textsubscript{2} storage solutions for distributed emitters, particularly in pilot-scale projects. However, their limited volumes, extensive alteration, low permeability, and insufficient site-specific data hinder scalability relative to Poland’s annual CO\textsubscript{2} emissions. Detailed site characterization and reactive transport modeling are essential for establishing a scientific foundation and addressing technical challenges.

\begin{table}[H]
\centering
\caption{\small Carbon Mineralization Potential (SRL 1) in Polish Mafic and Ultramafic Rocks. The numbers and identified prospects represent a high-level evaluation of theoretical potential and require careful further assessment to unlock the carbon mineralization potential in these formations.}
\label{tab:carbon_mineralization_summary}
\color{black}
\arrayrulecolor{black}
\footnotesize
\begin{tabular}{p{3cm}p{2.5cm}p{2cm}p{2cm}p{2.5cm}p{2.5cm}}
\toprule
\textbf{Formation} & \textbf{Location} & \textbf{Rock Type} & \textbf{Extent (km²)} & \textbf{CO\textsubscript{2} Capacity (Mt, SRL 1)} & \textbf{Key Constraints} \\
\midrule
\textbf{Ślęża Ophiolite} & Sudetes Mts. & Serpentinites, gabbros & 20 & 1-10 & Low permeability, serpentinization \\
\textbf{Nowa Ruda} & Sudetes Mts. & Serpentinites, gabbros & <10 & 5-20 & Small volume, fragmentation \\
\textbf{Holy Cross Mts.} & Central Poland & Basalts, diabases & Variable & 5-20 & Limited extent, shallow depth \\
\textbf{Gorzów Block} & NW Poland & Andesite-basalts & Patchy & Limited & Deep burial, metamorphism \\
\textbf{Kaczawa/Złotoryja} & SW Poland & Basaltic lavas & Limited & Limited & Small volume \\
\midrule
\textbf{Total Estimated} & \textbf{All regions} & \textbf{Mixed} & \textbf{<50} & \textbf{10-50} & \textbf{Scale limitations} \\
\bottomrule
\end{tabular}
\end{table}

\section{Discussion}
\label{sec:discussion}

\subsection{Geological Sequestration Assessment}

Poland's geological diversity provides substantial CO\textsubscript{2} storage capacity that positions the country as both a domestic solution for decarbonization and a potential regional CCS hub. Poland's geographical position in Central Europe provides strategic advantages for CCS deployment, with major industrial centers distributed across distinct geological regions: Upper Silesia in the south (responsible for ~40\% of national emissions), the Warsaw-Łódź corridor in central Poland, Wielkopolska in the west-central region around Poznań, and the Pomeranian coast along the Baltic Sea. This distribution aligns with Poland's three major storage regions—the Polish Lowlands, Baltic Basin, and Carpathian Foredeep—enabling proximity-based storage solutions while offering access to international North Sea storage infrastructure via Baltic shipping routes.

{\textbf{Storage capacity and regional distribution.}} Poland's diverse geological landscape offers formations suitable for large-scale CO\textsubscript{2} sequestration, with total theoretical capacity (SRL 1-3) estimated at 8.9-15 Gt distributed across three major geological regions (Figs. \ref{fig:F7_basins}-\ref{fig:F8_storage}): the Polish Lowlands (encompassing the Warsaw-Łódź industrial corridor and Wielkopolska region), the Baltic Basin (offshore Pomeranian coast), and the Carpathian Foredeep (extending from Kraków eastward) \cite{tarkowski2010potencjalne,wojcicki2014assessment, Ringstad2023CCUSZEN}. These regions feature sedimentary formations with favorable reservoir characteristics and thick caprock layers of mudstone/shale or anhydrite \cite{tarkowski2011petrophysical}.

The Polish Lowlands Basin, covering 72\% of Poland's territory, accounts for approximately 85\% of the country's theoretical storage potential. Key sub-regions include the Bełchatów Zone with Lower Jurassic anticlinal traps, the Mazovia Trough featuring multi-reservoir Jurassic-Cretaceous systems near Warsaw-Płock, the Poznań area within the Fore-Sudetic Monocline housing Rotliegend volcaniclastics, and northwestern Poland's Choszczno-Suliszewo Triassic-Jurassic structures. The Carpathian Foredeep offers dual storage mechanisms in Badenian sandstones and Carboniferous-Devonian carbonates, while the Baltic Basin provides cross-border potential through Cambrian Deimena Formation sands within Poland's exclusive economic zone.

\textbf{Storage architecture and strategic options.} Storage potential is dominated by two primary reservoir types:

\begin{itemize}
    \item \textbf{Deep Saline Aquifers:} Representing 90–93\% of Poland's total geological storage potential, positioned at depths ensuring supercritical CO\textsubscript{2} conditions. Their abundance and capacity for large storage volumes, combined with widespread availability and proximity to emitter sites (Fig. \ref{fig:F8_storage}) position them as key domestic solutions for CO\textsubscript{2} storage.

    \item \textbf{Depleted Oil and Gas Reservoirs:} Contributing 7-10\% of total potential, these reservoirs leverage existing wells, subsurface data, and proven containment capabilities. They offer significant advantages in cost-effectiveness, deployment speed, and safety, though with smaller individual capacities than saline aquifers.
\end{itemize}

CO\textsubscript{2} storage in coal seams accounts for less than 1\% of potential capacity \cite{wojcicki2014assessment}. Similarly, carbon mineralization in mafic and ultramafic rocks offers permanent sequestration potential but remains theoretical due to limited research. Both coal seams and mineralization require detailed petrographic analyses, THMC modeling, and field-scale pilot tests to quantify storage efficiency.

\textbf{Site-specific opportunities and challenges.} Poland's storage portfolio showcases diverse opportunities with distinct strategic and technical profiles. Onshore saline aquifers, such as the Konary structure near Poznań in the Wielkopolska region, are within 100 km of several major emitters. Despite its proximity to industrial clusters and technical suitability, local governments in Wielkopolska opposed a 2023 injection permit, citing seismic concerns, despite PGI's probabilistic hazard assessment indicating minimal risk \cite{lubon2016effect}. The EU CCUS ZEN project independently selected this site among the most promising storage candidates, proposing storage in Konary and Kamionki aquifer structures serving four emitter subclusters (totaling 8.187 Mt/year emissions) via 4-38 km pipelines \cite{Gravaud2023, Shogenova2025_valuechain}. This case illustrates the challenge of bridging technical feasibility with social acceptance in populated regions.

The Baltic Basin offshore the Pomeranian coast is notable for its Cambrian and Devonian sandstone reservoirs. These structures offer estimated capacities of 3.8 Gt in saline aquifers alone \cite{nagy2022new, wojcicki2014assessment}, with the strategic advantage of 50 km proximity to Gdańsk's planned ECO\textsubscript{2}CEE terminal, minimizing transport costs. However, subsea pipeline development through ecologically sensitive Baltic habitats faces environmental opposition, while the Helsinki Convention creates regulatory complexity requiring consensus among all Baltic Sea states for sub-seabed activities.

{\textbf{Technical assessment requirements.} Site characterization remains essential given Poland's diverse geological settings and significant variations in reservoir properties. Critical assessments must encompass stratigraphic architecture, reservoir quality distributions, caprock integrity, THMC evaluations, and fault stability. 

Recent research highlights caprock complexity challenges. Early assessments presumed thick mudstone or claystone caprocks guaranteed effective sealing and structural integrity. However, even thick clay-rich formations exhibit petrophysical, physical, and geomechanical properties that vary by orders of magnitude \cite{nooraiepour2017experimental, nooraiepour2017compaction, nooraiepour2018rock, nooraiepour2019permeability, nooraiepour2022clay}, significantly impacting leakage risk assessments and pressure management strategies.

Thermo-Hydro-Mechanical-Chemical (THMC) processes present critical technical challenges that necessitate an integrated assessment. In saline aquifers with elevated salinity levels (>10,000 mg/L) both onshore (Polish Lowland Basin) and offshore (Baltic Sea), CO\textsubscript{2} injection can trigger salt precipitation through CO\textsubscript{2}-induced evaporation-precipitation processes \cite{nooraiepour2025three,nooraiepour2018effect,dkabrowski2025surface, masoudi2021pore}, potentially clogging pore spaces, reducing injectivity, and compromising containment integrity \cite{nooraiepour2025potential,masoudi2024mineral, nooraiepour2018salt, aminzadeh2025ultrasonic}.

{\textbf{Integration and deployment pathways.} Poland's storage strategy requires balancing onshore accessibility with offshore capacity while addressing distinct regulatory and social challenges. Onshore reservoirs offer proximity to industrial emission clusters in Łódź and Warsaw, minimizing transport costs and infrastructure complexity \cite{marek2011geological}. However, public opposition rooted in perceived risks has slowed permitting processes. Offshore reservoirs in the Baltic Sea circumvent land-use disputes but face distinct challenges. The Baltic Cambrian Formation (80–150 km offshore from the Pomeranian coast) provides significant capacity with Zechstein evaporites as caprock units. However, offshore development encounters regulatory constraints under the Helsinki Convention requiring unanimous Baltic Sea state approval, plus 30-50\% higher infrastructure costs driven by subsea wellhead systems and corrosion-resistant materials \cite{kuczynski2018economic,volpi2015evaluation}. These cost premiums reflect the technical complexity of marine CO\textsubscript{2} injection and the specialized equipment required for Baltic Sea conditions.

International offshore storage via North Sea projects like Northern Lights provides additional strategic flexibility \cite{nooraiepour2025norwegian, furre2019building}. However, transporting CO\textsubscript{2} from Polish industrial centers to North Sea sites requires robust infrastructure development, including CO\textsubscript{2} purification facilities to meet stringent quality standards (99.9\% purity) \cite{bielka2023co2,bielka2024risks}, potentially adding €25-30 per tonne and creating administrative complexity across multiple jurisdictions \cite{kuczynski2018economic,kuczynski2014application}.

{\textbf{Assessment maturity and advancement pathway.} The provisional storage capacity estimates presented in this synthesis review underscore Poland's subsurface geological potential but carry significant uncertainty over wide probability percentiles due to reliance on static theoretical methods. Advanced dynamic reservoir simulations, informed by geophysical, geological, geochemical, geomechanical, and hydromechanical studies, combined with drilling and testing, are crucial for validating storage structures, assessing rock behavior during injection, and ensuring operational safety while progressing systematically through Storage Readiness Levels (Figure \ref{fig:F5_srl}). These efforts must be underpinned by enhanced research and development (R\&D) and strengthened academia-industry partnerships, as outlined in \cite{nooraiepour2025norwegian}.

Securing access to external CO\textsubscript{2} storage reservoirs, such as those in the North Sea \cite{nooraiepour2025norwegian, Lothe2019capacity}, appears to be critical to accommodate emissions from Poland's high-priority industrial sectors. This strategy requires robust governmental and commercial agreements to enable cross-border storage solutions. Concurrently, preparatory and pilot projects must commence immediately to lay the groundwork for large-scale deployment. The question of whether depleted hydrocarbon reservoirs, offering higher storage readiness levels, should be prioritized over saline aquifers or hybrid systems requires detailed techno-economic analyses and comprehensive subsurface characterization.

Poland's optimal storage strategy requires a diversified approach leveraging multiple geological options while progressing systematically through Storage Readiness Levels. A hybrid deployment model combining onshore proximity advantages with offshore capacity potential, supported by robust scientific research, engineering solutions, and proactive policy frameworks, will be essential for achieving Poland's decarbonization objectives while maintaining alignment with European climate neutrality commitments.

\subsection{Integration within the Value Chain}

Large-scale CO\textsubscript{2} storage deployment in Poland requires robust integration across the five-stage CCS value chain: capture, compression and purification, transport, utilization, and storage. This holistic approach transforms CO\textsubscript{2} from an industrial byproduct into a managed resource, aligning infrastructure, policy, and market mechanisms to build a resilient and scalable system critical to Poland's Energy Policy 2040. The success of this integration depends on coordinated development across all value chain components, where delays or inadequacies in any single stage can constrain the entire system's effectiveness.

Strategic site selection underpins effective integration by balancing geological suitability, proximity to emission sources, and public acceptance \cite{Wójcicki2008,Labus20152493,lewandowska2024screening}. Poland's distributed CO\textsubscript{2} storage capacity necessitates a regionally tailored "cluster-and-hub" model, where geographically concentrated industrial emissions are captured and aggregated before transport to suitable storage sites. Dense industrial zones like Upper Silesia (~40\% of Poland's emissions) \textcolor{blue}{can} connect directly to local storage via short pipelines, while smaller emitters feed into regional hubs for aggregation and large-scale sequestration \cite{IEA_Poland_Emissions_2023}. Gdańsk represents a strategic location that could serve as a major domestic and international CO\textsubscript{2} transit hub, facilitating both domestic applications and offshore storage in the Baltic and North Seas.

Poland's transport infrastructure strategy encompasses multiple modalities with distinct economic profiles, each addressing different scales and geographical constraints. The proposed pipeline connecting Upper Silesia to Baltic storage sites represents the most ambitious initiative, targeting 20 Mtpa capacity by 2040. However, construction challenges in mountainous regions and land acquisition complexities necessitate phased implementation prioritizing high-emission industrial clusters.

Alternative transport solutions address smaller-scale emitters where direct pipeline access proves uneconomical. Liquefied CO\textsubscript{2} rail transport, despite higher costs (€15–20/tonne versus €8–12/tonne for pipelines), offers flexibility for industrial facilities lacking pipeline connectivity, though it remains constrained by rail network capacity and terminal infrastructure. The proposed Oder River barge corridor, inspired by the Rhine CCUS network, could transport CO\textsubscript{2} from Wrocław cement plants to storage hubs, though seasonal water level fluctuations pose reliability challenges.

Maritime shipping emerges as a strategic solution for international CO\textsubscript{2} movement, providing Poland with access to established North Sea storage infrastructure. The ECO\textsubscript{2}CEE Project proposes a Gdańsk terminal collecting emissions from Poland, Czechia, and Slovakia for shipment to Norway's Northern Lights facility. While maritime transport provides flexibility and mitigates domestic storage delays, it requires 99.9\% CO\textsubscript{2} purity (adding €25–30/tonne) and raises energy sovereignty concerns regarding reliance on foreign storage sites.

Monitoring, Risk Management, and Operational Integration. Rigorous monitoring and risk management systems must be integrated across the entire value chain to ensure storage security and environmental integrity through advanced reservoir simulation, real-time geophysical monitoring, and geochemical assessments that track CO\textsubscript{2} plume movement and potential leakage risks \cite{Kabir2024,Bartela2014513,kaiser2014development}. These systems are critical for regulatory compliance and public trust, particularly given Poland's seismic and groundwater concerns.

Carbon Utilization and Economic Viability Enhancement.}} CO\textsubscript{2} utilization (CCU) offers complementary pathways that enhance CCUS economic viability by converting emissions into value-added products, creating additional revenue streams that can improve overall project economics. Polish initiatives are exploring synthetic fuels, polymers, and mineralized building materials, including CO\textsubscript{2}-curing concrete technologies, which are under evaluation by construction consortia. Chemical sector applications such as urea and methanol production present near-term opportunities within industrial clusters where hydrogen and CO\textsubscript{2} streams can be co-located. However, CCU faces economic and policy hurdles including high energy requirements, low product margins, and limited regulatory recognition. Despite these challenges, CCU provides strategic value as a bridge technology that can reduce net costs and promote early-stage adoption in hard-to-abate sectors, while permanent geological storage remains the primary pathway for large-scale emission reductions.

\subsection{Infrastructure, Regulatory, and Financial Considerations}

Establishing Poland's CO\textsubscript{2} storage network requires integrated infrastructure development, regulatory clarity, and financial mechanisms to drive decarbonization \cite{Zhang2021}. However, the interdependencies among these three pillars create complex coordination challenges that distinguish successful CCS deployment from technical feasibility studies. Without coordinated progress across these three pillars, even technically feasible storage projects risk delays or failure due to permitting uncertainties, investor hesitation, and infrastructure gaps. International experience demonstrates that sequential rather than parallel development of these elements often leads to project failure, as regulatory delays undermine infrastructure investments.

{\textbf{Infrastructure development: Balancing deployment with pragmatism.} Poland's CCUS commitment necessitates comprehensive CO\textsubscript{2} transport and storage infrastructure, balancing new construction with strategic repurposing of existing assets. Retrofitting natural gas pipelines for CO\textsubscript{2} transport offers significant cost advantages, requiring only 1–10\% of new infrastructure capital expenditure \cite{Isoli2025}. However, CO\textsubscript{2}'s distinct thermophysical properties—higher density and lower viscosity in supercritical conditions—demand modifications in pipeline materials, flow control systems, and compression requirements. This retrofit approach, while economically attractive, faces technical limitations in high-pressure applications and requires careful assessment against aging gas infrastructure, much of which may lack the specifications necessary for safe CO\textsubscript{2} transport.

Critical technical considerations include corrosion mitigation, as water impurities in CO\textsubscript{2} streams form carbonic acid that accelerates material degradation. Mitigation strategies encompass corrosion-resistant coatings, internal linings, and pressure-regulated control systems to prevent material fatigue and leakage \cite{Pal2023,Śliwińska2022}. Real-time monitoring systems, integrating advanced seismic sensors, pressure gauges, and continuous well-logging technologies, enable the early detection of anomalies. Digital twin modeling enhances performance forecasting and regulatory compliance.

Poland's flagship infrastructure initiative, the ECO2CEE project, establishes an open-access, multi-modal CO\textsubscript{2} terminal at Gdańsk Port, connecting Polish industrial emitters with offshore storage sites via the European CO\textsubscript{2} transport network. Scheduled for mid-2026 operations, the terminal targets 3 million tonnes CO\textsubscript{2} annually with planned capacity expansion \cite{orlen2025ccs}. However, this timeline appears optimistic compared to similar projects in Rotterdam and Antwerp, which experienced 2-3 year delays due to permitting complexities and technical integration challenges.

Given the capital-intensive nature of extensive pipeline networks, Poland envisions state-owned entities managing national CO\textsubscript{2} transportation networks while ensuring open access and standardized operations. Multi-modal transport solutions—encompassing pipelines, rail, road, and maritime options—provide flexibility for linking diverse emission sources with storage sites. The proposed "direct CO\textsubscript{2} pipelines" legal category aims to streamline the connections between capture installations and storage sites, thereby accelerating the development of point-to-point infrastructure.

Safety considerations remain paramount, given the public's acceptance requirements and operational security. Poland prioritizes stringent safety standards, transparent system design, and extensive monitoring protocols for large-scale storage projects. Technical regulations specify pipeline material requirements, operating conditions, and connection protocols while pilot storage projects demonstrate secure underground injection to build community confidence. The government is developing a comprehensive risk assessment framework to evaluate and mitigate transport and sequestration hazards, ensuring minimal operational risk as infrastructure scales.

{\textbf{Regulatory evolution: From fragmentation to integration.} Poland's regulatory landscape has undergone reforms to establish more supportive legal frameworks that ensure environmental compliance and safety standards. The 2023 amendments to the Geological and Mining Law enhanced the legal foundation for onshore CO\textsubscript{2} storage, addressing some of the previous regulatory gaps that constrained large-scale deployment \cite{Ga̧siorowska201043,DzU2023_XXX}.

Key regulatory advances include precise definitions of liability responsibility, ensuring operator accountability throughout both operational and post-closure phases. The amendments establish comprehensive long-term monitoring obligations requiring continuous assessment of CO\textsubscript{2} plume behavior, pressure fluctuations, and leakage risks. However, the 20-year minimum monitoring requirement creates significant long-term liability exposure that may deter private investment. These reforms should eventually lead to permitting onshore CO\textsubscript{2} storage and introduce incentives for small-scale projects, while simplifying regulatory processes by replacing exploration licenses with approvals for geological work projects.

The regulatory framework faces critical tests in balancing industrial development needs with environmental protection, particularly regarding groundwater resources and seismic risks. Poland's dense population and intensive groundwater utilization create regulatory complexity, requiring sophisticated risk assessment protocols and robust stakeholder consultation processes. 

Regulatory clarity has a direct impact on attracting investment and promoting industrial stakeholder participation. Unclear liability frameworks and ambiguous closure requirements have historically discouraged private-sector participation \cite{Muslemani2020}. Streamlined permitting processes and well-defined remediation protocols reduce investor uncertainties while ensuring stringent operational standards.

Public acceptance challenges require transparent regulatory mechanisms, enhanced engagement strategies, and accessible risk communication. European evidence demonstrates that public trust increases significantly when regulatory frameworks include independent monitoring and well-defined emergency response plans \cite{Wesche202348,Nguyen2022211}. Poland's evolving framework emphasizes transparency and community engagement to address societal concerns regarding storage safety, groundwater protection, and induced seismicity.

{\textbf{Financial architecture: Risk distribution and investment incentives.} Large-scale CO\textsubscript{2} storage deployment requires substantial financial investment across multiple funding sources. Poland allocates significant resources through the Energy Transformation Fund to support CCUS infrastructure, research, and risk management while enhancing industrial competitiveness under the EU Emissions Trading System \cite{Bacci2023523,Strojny2024}. This funding approach may create fiscal sustainability concerns, particularly given Poland's existing energy transition commitments and EU recovery fund obligations.

The economic case for CCUS in Poland hinges on carbon price trajectories and industrial competitiveness considerations. Current EU ETS prices of €60-80 per tonne CO\textsubscript{2} make CCUS marginally viable for some applications, but price volatility creates investment uncertainty. Poland's heavy reliance on carbon-intensive industries makes CCUS economically essential for maintaining industrial competitiveness, yet this same dependence limits fiscal capacity for public investment.

CCUS economic benefits extend beyond emission reductions through compliance cost reduction and CO\textsubscript{2}-enhanced oil recovery opportunities that extend field life and generate revenue \cite{Mathisen20176721}. Public-private partnerships and EU structural funds play crucial roles in investment de-risking and scalability acceleration, while tax incentives similar to the US Section 45Q program could stimulate private sector involvement.

European funding mechanisms, particularly the EU Innovation Fund, support demonstration and pilot projects by improving the economic viability of marginal initiatives. The Gdańsk terminal's EU Project of Common Interest recognition exemplifies how European funding facilitates critical infrastructure development. Past experiences, including the Bełchatów project's insufficient EU funding without domestic guarantees, highlight the importance of blended financing and government support.

Poland's CCUS investment strategy centers on public-private partnerships, recognizing substantial capital requirements. Public funding or EU financing may cover initial capital costs for capture installations in industries that are urgently decarbonizing, while private companies manage operational expenditures. For transport and storage infrastructure—considered natural monopolies—substantial state involvement and EU financing are anticipated.

Innovative financial instruments under consideration include Carbon Contracts for Difference (CCfDs) offering operators guaranteed minimum prices for CO\textsubscript{2} emission reductions. Government compensation when carbon market prices fall below agreed thresholds makes carbon capture economically viable while incentivizing early project development.

International collaborations enhance financing capabilities through initiatives such as cooperation with Norway's Northern Lights project, under the EU-funded ACCSESS program, which explores industrial CO\textsubscript{2} transport from Polish facilities to North Sea storage locations. Regional discussions on Central European CO\textsubscript{2} transport pipeline networks demonstrate how multinational partnerships attract broader investment, pool financial resources, and integrate national infrastructure with regional and European networks \cite{Zhang2021}. Poland's participation in regional initiatives, including the Baltic Sea CCUS network, enables collaborative financing and integration with European carbon storage centers. The proposed Gdańsk export terminal exemplifies how cross-border partnerships enhance economic feasibility while supporting Poland's transition to large-scale commercial CCUS deployment.

\subsection{Measurement, Monitoring and Verification (MMV)}

Poland's geological CO\textsubscript{2} storage implementation requires comprehensive MMV frameworks to mitigate risks, ensure EU regulatory compliance, and build public trust through transparent oversight \cite{Larkin2019286}. As an emerging CCS market, Poland must develop tailored monitoring strategies addressing its diverse storage portfolio while managing cost-efficiency and technological integration challenges.

Poland's varied geological storage candidates necessitate differentiated MMV approaches optimized for specific environmental and technical contexts. Onshore storage sites, particularly those near populated areas, employ non-invasive monitoring techniques including satellite-based interferometric synthetic aperture radar (InSAR) for detecting subtle surface deformations and time-lapse seismic surveys for tracking CO\textsubscript{2} plume migration \cite{Chałupnik20143}. These methods provide continuous surveillance while minimizing surface disruption and addressing community concerns about storage safety.

Offshore storage projects in ecologically sensitive Baltic Sea environments rely on autonomous underwater vehicles (AUVs) equipped with real-time geochemical sensors and sub-bottom profilers to assess storage integrity and detect potential leaks \cite{Kirchsteiger20081149}. For legacy hydrocarbon fields like the Barnówko-Mostno-Buszewo (BMB) site, electromagnetic field monitoring represents an innovative approach for assessing casing corrosion in existing wells \cite{Zonetti2023_monitoring}.

Advanced analytics integration enhances monitoring capabilities through AI-driven plume modeling and machine learning algorithms for spatial and temporal datasets and handling partial/ordinary differential equations \cite{nooraiepour2025traditional}. Drone-based soil CO\textsubscript{2} flux measurements and distributed sensor networks provide complementary data streams that, when integrated through centralized platforms, enable real-time monitoring and rapid response capabilities \cite{Leiss2019234}.

Poland's MMV systems must align with EU CO\textsubscript{2} storage directive requirements mandating annual leakage rates below 0.01\%, ensuring regulatory compliance while enabling proactive risk mitigation. The diverse geological storage portfolio presents distinct monitoring challenges and cost implications. Legacy sites such as BMB field benefit from preexisting well logs and seismic data, potentially reducing MMV costs, while greenfield sites in the Carpathian Foredeep require comprehensive preinjection characterization including 3D seismic mapping and noble gas isotope profiling to establish accurate baseline conditions.

The Polish Geological Institute estimates MMV systems account for 15–20\% of total storage costs, emphasizing the need for cost-efficient sensor technology innovations and data processing optimization \cite{PGI2025}. This cost consideration drives the development of hybrid monitoring approaches that balance technical performance with economic viability while maintaining regulatory compliance.

Poland's evolving regulatory landscape positions robust MMV frameworks as cornerstones of long-term storage security. Recent regulatory advances indicate a shift toward hybrid MMV networks combining state-managed sensors with independent third-party audits \cite{Ga̧siorowska201043}. This redundant oversight structure strengthens EU standards compliance while enhancing public confidence through transparent and accountable monitoring.

The MMV framework's role extends beyond technical monitoring to encompass risk communication and stakeholder engagement. Transparent data sharing, accessible reporting mechanisms, and community involvement in monitoring design help address public concerns about storage safety and environmental protection. Real-time data availability and independent verification processes establish the social license necessary for the successful deployment of CCS in Poland's diverse geological and social contexts.

Poland's MMV development emphasizes adaptability and scalability to accommodate future storage expansion while maintaining operational efficiency and public trust. The integration of emerging technologies, cost-effective monitoring solutions, and transparent governance structures positions Poland to establish a robust MMV framework supporting both domestic storage development and potential regional CCS hub capabilities.

\subsection{Public Engagement and Social Acceptance}

The successful deployment of CCUS technologies in emerging regions is fundamentally dependent on fostering public trust, social acceptance, and effective stakeholder engagement strategies that must be tailored to local contexts and socio-political landscapes. Figure \ref{fig:F11_trustpathway} illustrates a comprehensive framework for building public support, demonstrating that beyond technical and economic feasibility, securing societal acceptance represents a critical determinant of CCS project viability. Emerging CCS regions, including countries with historical fossil fuel dependencies and those transitioning to new energy economies, face unique challenges in public engagement that require nuanced approaches to communication, participation, and trust-building. For countries pursuing CCS as part of their decarbonization strategies, alignment with international transparency standards and stakeholder participation frameworks becomes essential for building credible engagement processes \cite{wojakowski2024communicating,langhellesocial}.

Central to effective public engagement across lower-maturity CCS countries is the meaningful involvement of local stakeholders in decision-making processes, moving beyond traditional top-down approaches toward collaborative governance models (Fig. \ref{fig:F11_trustpathway}). International experience, including lessons from the EU FP7 SiteChar project and similar initiatives, demonstrates that proactive engagement strategies incorporating social site characterization and early public consultations significantly enhance community trust and project acceptance across diverse cultural and political contexts \cite{Brunsting2015767,wojakowski2022public}. Economic co-benefits and comprehensive risk communication emerge as universal factors shaping public perception, with communities consistently showing greater support for CCS projects when they perceive clear local advantages in terms of employment opportunities, energy security enhancement, and environmental protection. The Polish experience exemplifies this pattern, where transparent information dissemination and direct involvement of community representatives have been shown to improve public confidence in CCS technologies \cite{nooraiepour2025norwegian,wojakowski2024communicating,langhellesocial}.

Trust in governmental institutions represents a critical variable that differs significantly across emerging regions, requiring tailored approaches to leverage existing social capital while addressing institutional weaknesses. In contexts where communities exhibit higher trust in local authorities—as observed in Poland—governmental support combined with transparent communication can effectively facilitate public acceptance \cite{wojakowski2022public}. However, regardless of institutional trust levels, effective public participation strategies must transcend passive information dissemination to enable active community involvement, empowering stakeholders to influence project design and implementation decisions. Equally critical is the development of robust risk management frameworks that address community concerns through stringent safety protocols, continuous monitoring systems, and transparent reporting mechanisms. Such frameworks must demonstrate unwavering commitment to environmental protection and public safety while maintaining open channels for community feedback and oversight.

\begin{figure}
    \centering
    \includegraphics[width=0.75\linewidth]{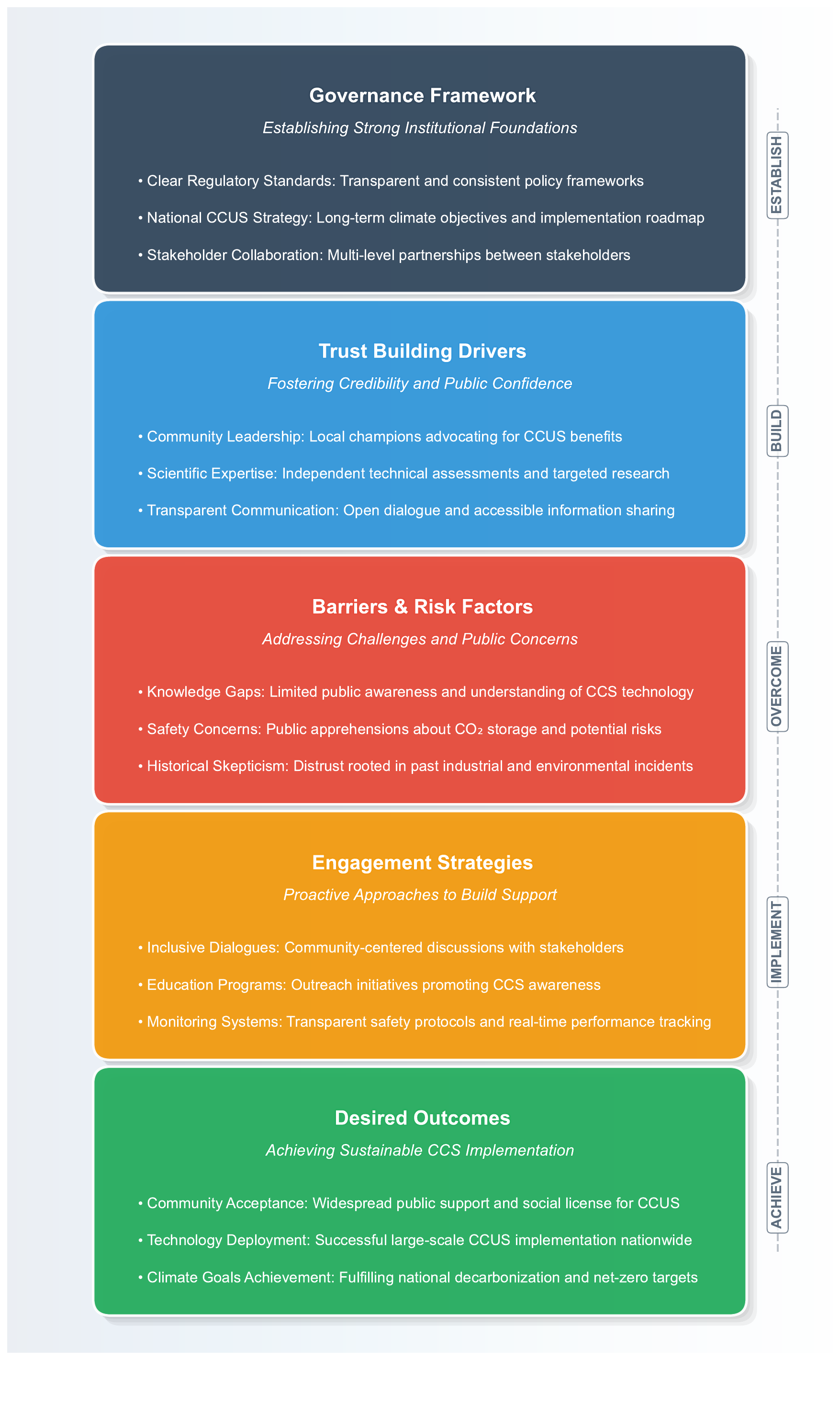}
    \caption{Framework for establishing public trust and stakeholder engagement in CCUS implementation for emerging CCS regions. The framework establishes a hierarchical pathway from governance foundations to desired outcomes. Strong governance structures establish transparent regulatory frameworks and national strategies, which are essential for policy consistency. Trust-building drivers focus on community leadership, scientific expertise, and transparent communication to build public confidence. The framework addresses key barriers, including knowledge gaps, safety concerns, and historical skepticism, through proactive risk management. Engagement strategies emphasize inclusive dialogues, educational outreach, transparent monitoring, and promoting economic benefits. The successful implementation of this framework leads to enhanced community acceptance, widespread technology deployment, and the achievement of national climate commitments, enabling a transition to a low-carbon society.}
    \label{fig:F11_trustpathway}
\end{figure}

Comparative analysis of public engagement experiences across established and emerging CCS regions reveals both universal principles and context-specific factors that influence social acceptance patterns. Cross-national studies spanning Belgium, France, Greece, Italy, Norway, Turkey, the United Kingdom, and other regions identify procedural justice, distributional equity, institutional trust, and public knowledge as fundamental determinants of CCS acceptance \cite{Karytsas2023}. While economic benefits serve as primary acceptance drivers in resource-dependent economies like Poland, international evidence suggests that fairness in decision-making processes and equitable risk-benefit distribution assume greater importance in regions with more diversified economies and established environmental governance frameworks. The United States experience further illustrates regional variation, where public support patterns differ markedly from European contexts, with notably reduced 'Not-in-My-Backyard' resistance, suggesting that cultural attitudes toward industrial infrastructure significantly influence acceptance dynamics \cite{Fikru2024425}.

Nevertheless, persistent concerns regarding technological uncertainty and potential prolongation of fossil fuel dependence appear consistently across diverse regional contexts, indicating that traditional information-deficit models of science communication prove insufficient for addressing deeply held skepticism about CCS technologies \cite{Upham20111359,Oltra2012227}. Addressing these concerns requires context-sensitive public engagement approaches that acknowledge local priorities, cultural values, and historical experiences, while providing accessible and relevant information that directly addresses community-specific concerns \cite{vonRothkirch2021,Stavrianakis2024}.

The evolution of public engagement strategies across emerging CCS regions demonstrates the critical importance of local stakeholder integration, with successful approaches varying significantly based on national governance structures, cultural contexts, and existing levels of institutional trust \cite{langhellesocial,wojakowski2024communicating,wojakowski2022public}. Regional variations in stakeholder participation effectiveness reflect broader differences in democratic traditions, civil society strength, and regulatory frameworks, requiring adaptive engagement models rather than standardized approaches \cite{Brunsting2015767,Karytsas2023}. Economic incentive structures and risk communication strategies must be tailored to local socio-economic conditions, with particular attention to communities experiencing the impacts of energy transition or industrial restructuring challenges.

Institutional trust levels vary considerably across emerging CCS regions, necessitating differentiated strategies that either leverage existing social capital or compensate for institutional weaknesses through enhanced transparency and participatory mechanisms. Where regulatory confidence remains limited, engagement strategies must prioritize procedural justice and equitable benefit-sharing to build legitimacy, while regions with stronger institutional foundations can focus more directly on technical communication and economic co-benefit development \cite{Oltra2012227,Brunsting2015767}.

Effective public engagement frameworks for emerging CCS regions must integrate transparent, evidence-based communication strategies that comprehensively address the entire CCS value chain while acknowledging local concerns and priorities. Multi-channel education campaigns that utilize both traditional media and digital platforms can help bridge knowledge gaps and counter misinformation, particularly in regions where public awareness of CCS technologies remains limited \cite{Kaiser2014267,Brunsting20137327}. Participatory decision-making assumes particular importance in economies undergoing energy transitions, where communities face uncertainties about industrial restructuring and the impacts on employment. Public consultation processes, community co-design initiatives, and collaborative planning mechanisms can effectively integrate local knowledge while addressing environmental and socio-economic concerns, simultaneously highlighting long-term benefits for energy security and green economy development \cite{wojakowski2022public,Rychlicki2016133}.

For emerging regions seeking to establish credible engagement processes, aligning with established international standards and best practices offers significant advantages in terms of technical learning and building stakeholder confidence. Regional frameworks such as the EU's Net Zero Industry Act and similar policy instruments provide valuable benchmarks for stakeholder participation and transparency requirements that can be adapted to local contexts while maintaining international credibility \cite{Karytsas2023}.

Ultimately, sustainable CCS deployment in emerging regions requires recognition that public acceptance represents not merely a social license to operate, but a fundamental prerequisite for long-term project viability and policy effectiveness (Figure \ref{fig:F11_trustpathway}). Well-designed engagement processes transform communities from passive recipients of industrial development into active partners in decarbonization initiatives, generating stronger regulatory support and enhanced investment attractiveness. In policy environments where public opinion increasingly influences infrastructure decisions, comprehensive engagement strategies combining scientific communication, participatory governance, robust risk management, and alignment with international frameworks become essential for realizing CCS potential across emerging regions. Success in overcoming public engagement challenges will ultimately determine whether carbon capture and storage technologies can achieve the scale and speed of deployment necessary for meeting global climate objectives while ensuring socially sustainable energy transitions.

\section*{Conclusions}

This study establishes a systematic framework for assessing geological CO\textsubscript{2} storage in emerging CCUS markets, addressing critical methodological gaps that constrain decarbonization strategies in lower-maturity regions worldwide. Using Poland as a representative case study, the research demonstrates how countries with carbon-intensive economies, evolving regulatory landscapes, and nascent relevant infrastructure can evaluate and develop their geological sequestration potential.

\textbf{Framework contribution and global applicability.} The integrated assessment methodology, combining the resource-reserve pyramid with Storage Readiness Level classifications, provides emerging CCS regions with standardized tools for progressing from theoretical capacity estimates to bankable storage projects. This framework addresses the challenges facing countries lacking extensive geological information and subsurface databases, established regulatory precedents, or mature stakeholder engagement protocols. The phase progression methodology, from regional screening through commercial deployment, offers adaptable templates for diverse geological, economic, and institutional contexts that pursue CCS integration.

\textbf{Poland's strategic position and sectoral implications.} As the EU's fourth-largest emitter, Poland's 8.9-15 Gt theoretical storage capacity positions the country as a potential regional CCS hub while addressing domestic decarbonization imperatives. The diversity of storage options—onshore Polish saline aquifers proximate to industrial clusters, offshore Baltic Sea formations offering cross-border potential, and repurposed hydrocarbon reservoirs providing near-term deployment opportunities—demonstrates how geological diversity can support flexible, risk-distributed storage strategies. Poland's model in the near future can set an example for navigating environmental regulations (Baltic Sea), cross-border storage negotiations, and industrial cluster integration, providing valuable precedents for other emerging markets facing similar multi-jurisdictional challenges.

\textbf{Critical barriers and systematic solutions.} Current storage assessments in Poland remain constrained at SRL 1-3 due to fundamental limitations: sparse subsurface characterization data, restricted industry-academia collaboration, and inadequate dynamic reservoir modeling capabilities. These deficiencies introduce uncertainty ranges exceeding business decision-making thresholds, highlighting the necessity for coordinated data acquisition campaigns, enhanced research partnerships, and standardized assessment protocols. The research identifies regulatory evolution as equally critical to technical advancement, requiring parallel development of legal frameworks, financial mechanisms, and stakeholder engagement strategies.

\textbf{Integration requirements and policy implications.} Successful CCS deployment in emerging markets demands simultaneous advancement across technical, regulatory, economic, and social dimensions rather than sequential development. The study's analysis of Poland's evolving policy landscape, from early EU directive transposition to recent industrial partnerships with prominent international actors, such as the Norwegian-Polish CCS network and the ORLEN-Equinor collaboration, demonstrates how coordinated value chain integration accelerates market maturation. The identification of public acceptance as a fundamental prerequisite—not merely a social license consideration—underscores the necessity for transparent governance frameworks and inclusive community engagement from project inception.

The framework developed through Poland's case study provides emerging CCS regions with evidence-based tools for systematic storage assessment while highlighting the multi-dimensional coordination required for successful deployment. These insights advance both the scientific understanding of geological sequestration assessment methodologies and provide practical guidance for policymakers navigating the complex intersection of climate commitments, energy security, and sustainable economic transition in regions where CCS represents both an environmental necessity and a strategic opportunity.

\section*{Appendix}

\footnotesize
\begin{longtable}{@{} l l l l r @{}}
\caption{\small A list of potential CO\textsubscript{2} storage sites identified in Poland (\cite{wojcicki2014assessment,Ringstad2023CCUSZEN} and references therin), categorized into "Deep Saline Aquifers" (DSA) and "Hydrocarbon Fields" (HF) (refer to Figure \ref{fig:F8_storage}). The storage sites are ranked by storage capacity from highest to lowest for each category. SRL stands for Storage Readiness Level (refer to Figure \ref{fig:F5_srl}). All listed items are daughter units, which are specific structural and stratigraphic traps contained within a storage unit. By definition, depleted hydrocarbon fields qualify as daughter units, utilizing their established geological structures to store CO\textsubscript{2} safely.} \\ \\
\toprule
Storage Name & Storage Type & Location & SRL & Capacity [Mt] \\
\midrule
\endfirsthead

\toprule
Storage Name & Storage Type & Location & SRL & Capacity [Mt] \\
\midrule
\endhead

\midrule
\multicolumn{5}{r}{\textit{Continued on next page}} \\
\endfoot

\bottomrule
\endlastfoot

Gostynin          & DSA                 & Onshore & 2 & 514 \\
Suliszewo        & DSA                 & Onshore & 2 & 509 \\
Wojszyce         & DSA                 & Onshore & 3 & 342 \\
Wyszogród        & DSA                 & Onshore & 2 & 315 \\
Konary J         & DSA                 & Onshore & 2 & 282 \\
Jeżów T          & DSA                 & Onshore & 2 & 277 \\
Rokita           & DSA                 & Onshore & 2 & 264 \\
Debrzno          & DSA                 & Onshore & 2 & 246 \\
Huta Szklana     & DSA                 & Onshore & 2 & 224 \\
Sochaczew J      & DSA                 & Onshore & 2 & 222 \\
Sierpc K         & DSA                 & Onshore & 2 & 212 \\
Sochaczew K      & DSA                 & Onshore & 2 & 206 \\
Choszczno        & DSA                 & Onshore & 2 & 205 \\
Wierzchowo       & DSA                 & Onshore & 2 & 194 \\
Bielsk-Bodzanów  & DSA                 & Onshore & 2 & 194 \\
Konary T         & DSA                 & Onshore & 2 & 182 \\
Jeżów J          & DSA                 & Onshore & 2 & 166 \\
Niecka Poznańska (G-U-B-P) & DSA      & Onshore & 2 & 165 \\
Grudziądz        & DSA                 & Onshore & 2 & 157 \\
Sierpc J         & DSA                 & Onshore & 2 & 152 \\
Kamionki J       & DSA                 & Onshore & 2 & 149 \\
Marianowo J\&T  & DSA                 & Onshore & 3 & 129 \\
Budziszewice-Zaosie & DSA              & Onshore & 2 & 107 \\
Radnica          & DSA                 & Onshore & 2 & 87 \\
Dzierżanowo      & DSA                 & Onshore & 2 & 84 \\
Lutomiersk       & DSA                 & Onshore & 2 & 78 \\
Trześniew        & DSA                 & Onshore & 2 & 78 \\
Kliczków J       & DSA                 & Onshore & 2 & 77 \\
Bodzanów        & DSA                 & Onshore & 2 & 77 \\
Kowalowo        & DSA                 & Onshore & 2 & 76 \\
Husów-Albigowa-Krasne & DSA           & Onshore & 3 & 63 \\
Bielsk          & DSA                 & Onshore & 2 & 63 \\
Trzebież         & DSA                 & Onshore & 2 & 61 \\
Turek           & DSA                 & Onshore & 2 & 61 \\
Wartkowice      & DSA                 & Onshore & 2 & 53 \\
Szubin          & DSA                 & Onshore & 2 & 47 \\
Tuszyn          & DSA                 & Onshore & 2 & 41 \\
Radlin          & DSA                 & Onshore & 3 & 17 \\
Zat Gdowska     & DSA                 & Onshore & 2 & 15 \\
Skoczów-Czechowice & DSA              & Onshore & 2 & 14 \\
Oświno K        & DSA                 & Onshore & 2 & 13 \\
Grobla          & DSA                 & Onshore & 2 & 13 \\
Żyrów           & DSA                 & Onshore & 2 & 11 \\

J\_15\_POL\_K   & DSA                 & Offshore & 2 & 102 \\
MBu\_28\_POL\_K & DSA                 & Offshore & 2 & 52 \\
J\_06\_POL\_H   & DSA                 & Offshore & 2 & 48 \\
J\_14\_POL\_K   & DSA                 & Offshore & 2 & 39 \\
MBu\_23\_POL\_K & DSA                 & Offshore & 2 & 36 \\
MBu\_19\_POL\_K & DSA                 & Offshore & 2 & 27 \\

Przemyśl         & HF                  & Onshore & 3 & 370 \\
Żuchlów          & HF                  & Onshore & 3 & 138 \\
Brońsko          & HF                  & Onshore & 3 & 91 \\
Bogdaj-Uciechów  & HF                  & Onshore & 3 & 88 \\
Jarosław         & HF                  & Onshore & 3 & 36 \\
Kościan S        & HF                  & Onshore & 3 & 35 \\
BMB              & HF                  & Onshore & 3 & 30 \\
Paproć           & HF                  & Onshore & 3 & 27 \\
Wilków           & HF                  & Onshore & 3 & 23 \\
Tarnów (miocen)  & HF                  & Onshore & 3 & 15 \\
Czeszów          & HF                  & Onshore & 3 & 10 \\
Lubiatów         & HF                  & Onshore & 3 & 10 \\
Paproć W         & HF                  & Onshore & 3 & 10 \\
Grodzisk\_Wlkp.  & HF                  & Onshore & 3 & 9 \\
Lubaczów (J)     & HF                  & Onshore & 3 & 9 \\
Brzostowo        & HF                  & Onshore & 3 & 8 \\
Góra             & HF                  & Onshore & 3 & 8 \\
Łąkta            & HF                  & Onshore & 3 & 6 \\
Niepołomice      & HF                  & Onshore & 1 & 6 \\
Węglówka         & HF                  & Onshore & 3 & 4 \\
Kamień Pomorski   & HF                  & Onshore-Offshore & 3 & 3 \\
Buk              & HF                  & Onshore & 3 & 1 \\
Dzieduszyce      & HF                  & Onshore & 3 & 1 \\
Osobnica         & HF                  & Onshore & 3 & 1 \\
Radoszyn         & HF                  & Onshore & 3 & 1 \\
Wysoka Kamieńska  & HF                  & Onshore & 3 & 1 \\
Nosówka          & HF                  & Onshore & 3 & 0.5 \\

B 3              & HF                  & Offshore & 3 & 11 \\
B 8              & HF                  & Offshore & 3 & 8 \\
\label{table:storagelist}
\end{longtable}

\section*{Acknowledgments}

The authors acknowledge the support received from the "Norwegian-Polish CCS Network: Acceleration of Climate Change Mitigation Technologies Deployment" project, funded through the EEA and Norway Grants under the Green Transition Collaborations Programme (FWD-Green-11). We also thank the “Understanding Coupled Mineral Dissolution and Precipitation in Reactive Subsurface Environments” project, funded by the Norwegian Centennial Chair (NOCC).

\printbibliography

\end{document}